\RequirePackage[figuresright]{rotating}
\documentclass[useAMS,usenatbib]{mnras}

\usepackage{tabto}
\usepackage{float}
\usepackage{color}
\usepackage{amssymb} 
\usepackage{amsfonts,amsmath,amscd} 
\usepackage[subnum]{cases}
\usepackage{mathtools}
\usepackage{verbatim}
\usepackage{t1enc}
\usepackage[T1]{fontenc}
\usepackage{ae,aecompl}
\usepackage{graphics}
\usepackage{adjustbox}
\usepackage{caption}
\usepackage{eufrak} 
\usepackage{euscript} 
\usepackage[normalem]{ulem} 
\usepackage{makeidx} 
\usepackage[usenames,dvipsnames]{pstricks}
\usepackage{epsfig}
\usepackage{pst-grad} 
\usepackage{pst-plot} 
\usepackage{pstcol,times,pstricks,pst-node,pst-coil} 
\usepackage{multicol} 
\usepackage{multirow} 
\usepackage{indentfirst} 
\usepackage{natbib} 
\usepackage{setspace} 
\usepackage{threeparttable}

\newcommand{\specialcell}[2][c]{%
  \begin{tabular}[#1]{@{}c@{}}#2\end{tabular}}

\title[IBP for SC simulations]
{On the initial binary population for star cluster simulations}
\author[Belloni et al.]
{Diogo Belloni$^{1,2}$\thanks{E-mail: belloni@camk.edu.pl (DB)}, Abbas Askar$^1$, Mirek Giersz$^1$, Pavel Kroupa$^{3,4}$ and
\newauthor Helio J. Rocha-Pinto$^{5}$\\
$^{1}$ Nicolaus Copernicus Astronomical Centre, Polish Academy of Sciences, ul. Bartycka 18, PL-00-716 Warsaw, Poland \\
$^{2}$ CAPES Foundation, Ministry of Education of Brazil, DF 70040-020, Brasilia, Brazil \\
$^{3}$ Helmholtz-Institut f\"ur Strahlen- und Kernphysik, Nussallee 1416, D-53115 Bonn, Germany \\
$^{4}$ Charles University in Prague, Faculty of Mathematics and Physics, Astronomical Institute, V Holesovickach 2, CZ-180 00 Praha 8, Czech Republic \\
$^{5}$ Universidade Federal do Rio de Janeiro, Observat{\'o}rio do Valongo, Ladeira do Pedro Ant\^onio 43, 20080-090 Rio de Janeiro, Brazil}

\begin{document}

\date{Accepted... . Received... ; in original form...}

\pagerange{\pageref{firstpage}--\pageref{lastpage}} \pubyear{2017}

\maketitle

\label{firstpage}

\begin{abstract}

Colour-magnitude diagrams (CMDs) are powerful tools that might be used to infer stellar 
properties in globular clusters (GCs), for example, the binary fraction and their 
mass ratio ($q$) distribution. In the past few years, observations have 
revealed that q distributions of GC main-sequence (MS) binaries are generally flat, 
and a distribution characterized by a strong increase towards q $\approx$ 1
is not typical in GCs. In numerical simulations of GC evolution with the initial binary population (IBP)
described by Kroupa, synthetic CMD colour distributions exhibit 
a peak associated with binaries that have q $\approx$ 1. While
the Kroupa IBP reproduces binary properties in star-forming regions, clusters and the Galactic field, 
the peak in the q distribution towards q $\approx$ 1 observed for GC simulations is not consistent 
with distributions derived from observations.
The objective of this paper is to refine and further improve the physical formulation of pre-main-sequence 
eigenevolution proposed by Kroupa in order to achieve CMD colour distributions
of simulated GC models similar to those observed in real GCs, and to get a
similarly good agreement with binary properties for late-type binaries in the Galactic field.
We present in this paper a modified Kroupa IBP,  in which early-type stars follow observational distributions, 
and late-type stars are generated according to slightly modified pre-main-sequence eigenevolution prescriptions. 
Our modifications not only lead to a qualitatively good agreement with respect to long-term 
observations of late-type binaries in the Galactic field, but also resolve the above-mentioned 
problem related to binary distributions in GC models.

\end{abstract}


\section{INTRODUCTION}
\label{introduction}

The {\it initial binary population} (IBP) corresponds to initial
properties of binaries in star clusters, which follow particular
distributions of their parameters, i.e.primary mass, period,
eccentricity, and mass ratio. The concept of an IBP is important
because the full numerical solution  to the problem of cloud collapse
leading to star or binary formation is not viable \citep[e.g.][]{Kroupa_2011}, but a
description of initial populations is nevertheless needed for a wide
variety of astrophysical problems -- e.g. star cluster modelling,
stellar population synthesis and dynamical population synthesis
\citep[e.g.][]{Kouwenhoven_2009}.

It has been proposed that the field population of stars comes from the 
dissolution of star clusters \citep[e.g.][]{Lada_2003} after the expulsion 
of the residual gas in the star formation process. Clustered star formation 
and dissolution of these embedded clusters via gas expulsion is the dominant 
process that populates the field with stars 
\citep[e.g.][]{Lada_2003,Bressert_2010,Lada_2010,Kroupa_2011,Marks_2011b}. 
Additionally, it has been argued that the outcome of star-formation processes 
depends only weakly on the physical conditions of the molecular cloud 
\citep[e.g.][and references therein]{Kroupa_2011}. This is intrinsically 
associated with the {\it IBP universality hypothesis} which corresponds to 
an environment-independent star formation process 
\citep[][see their section 5.2.1]{Marks_2015}. 

Alternatively, some authors suggest that the star-formation process is not 
universal \citep[e.g.][]{King_2012,Parker_2014}. 
For example, \citet[][]{King_2012}, 
on the basis of the comparison between the properties of close binary systems in seven 
young regions and in the field, conclude that the origin of multiplicity is not universal.
In a similar way, \citet{Parker_2014} 
argue that the binary population in the field is indicative of the primordial 
binary population in star-forming regions, at least for systems with primary 
masses in the range 0.02 -- 3.0 M$_\odot$, based on comparisons between $N$-body 
simulations and binary properties in the Galactic field.

Concerning these two competing scenarios, we emphasize that both have
the potential of explaining the observed data. However, as discussed
in \citet[][]{Marks_2015},
the hypothesis of the cluster origin of the galactic binary population 
in connection with the universality hypothesis is the only currently 
available approach that allows binary populations to be predicted 
in the Milky Way and other galaxies \citep{Marks_2011b}, and for such a 
reason this scenario is assumed in this work.

A promising candidate for the universal IBP has been inferred from
observational data \citep{Kroupa_1995b,Kroupa_2013}, hereafter called
the Kroupa IBP. It has been tested against observations, and has 
successfully explained the observational features of young clusters, 
associations and Galactic field late-type binaries 
\citep[e.g.][]{Kroupa_2011,Marks_2012}. In addition, 
\citet{Belloni_2017a} found that models that follow the Kroupa IBP 
show good agreement with the observed Galactic cataclysmic variable 
white dwarf mass distribution. Direct imaging studies indeed reveal 
that most stars form in wide binaries \citep{Sadavoy_2017}.

Regarding the binary fraction in GCs, an efficient way of detecting
the binary content is the search for binary main-sequences (MS) in
colour-magnitude diagrams (CMDs) \citep[e.g.][]{Rubenstein_1997}.
\citet{Sollima_2007} were the first to investigate binaries in a
reasonably large sample of 13 GCs, and found 
that all the analysed globular clusters contain a minimum binary fraction 
larger than 6 per cent within the core radius and that the global binary fractions 
lie between 10 to 50 per cent (depending on the cluster). 
\citet{Milone_2012}, who extended the sample
and analysed MS binary populations in a sample of 59 GC CMDs,
found for the first time an anti-correlation 
between the binary fraction and the total cluster mass.

Considering that, \citet{Leigh_2015} investigated
whether or not the observed present-day distribution of Galactic GC
binary fractions can be reproduced assuming the Kroupa IBP 
(i.e. with a significant fraction of soft binaries).
These authors showed that high initial binary fractions with a 
significant soft component (Kroupa IBP) combined with high initial densities can 
reproduce the observed anti-correlation between the binary fraction 
(both inside and outside the half-mass radius) and the total cluster 
mass, which corroborates the idea that an environmental independent 
universal IBP might exist associated with the binary population in the Galactic field,
in associations, in young clusters and also in GCs.

Concerning the observed MS binary mass ratios in Galactic GCs, 
\citet{Milone_2012} found that the distribution of the mass ratio 
is generally flat (for $q>0.5$), with the exception of only
four GCs (E3, Terzan 7, NGC6366 and NGC6496). Since the MS binary mass ratio
distribution is associated with the colour distribution on the red side
of the fiducial MS, we should expect that such a 
distribution is also typically flat.

So far, no attempt to compare predicted and observed
GC MS binary distributions has been carried out, taking into account the Kroupa IBP. 
On this regard, we will show in this work (Section \ref{problem})
that, even though initial models following the Kroupa IBP predict present-day
binary fractions supported by observations, there is
a problem related to present-day MS binary distributions. 
In its original formalism, the Kroupa IBP contains a large
fraction of binaries with approximately equal masses (i.e. $q \approx
1$, where $q$ is the mass ratio). This causes a clearly visible binary
sequence close and above the MS turn-off in present-day GC CMD, and also colour
distributions (below the turn-off) characterized by a strong increase towards 
the right edge of the distribution. Such features are, however, not commonly 
observed in GCs \citep{Milone_2012} 

Rather than providing recipes for simulating present-day binary populations in GCs,
the objective of this paper is first to show that the Kroupa IBP 
predicts present-day MS binary distributions (via analysis
of present-day CMD colour distributions) which disagree with the observed ones, 
and second to develop a refined Kroupa IBP such that it 
can solve the above-mentioned problem with respect to GC MS binary
distributions. Our modifications not only 
solve the problem, but also provide
{
similar good
}
agreement with Galactic field 
late-type binaries. In addition, this corroborates even more the universality 
hypothesis and provides a step forward in better descriptions for energy and 
angular momentum redistribution processes during star formation.

The paper is structured as follows. In Section \ref{assumptions} we define 
the main concepts and assumptions used in this investigation. In Section 
\ref{original}, we introduce the Kroupa IBP and in Section \ref{problem}, 
we present the problem associated with the original prescription. 
The codes utilised in this investigation are described in Section \ref{codes},
and a refined version which solves the problem with MS binary distributions 
is developed in Section \ref{modified}. We show the main results 
with respect to binaries in the Galactic field and in GCs in 
Section \ref{results}. A summary of the main points of this investigation 
is provided in Section \ref{discussion}.


\section{DEFINITIONS AND ASSUMPTIONS}
\label{assumptions}

In order to clarify the terminology adopted in this work,
in this section we define the main concepts explored and
the main assumptions adopted. Note that the concept of IBP 
was already defined in Section \ref{introduction} and it will be 
skipped in this section.

\citet{Kroupa_1995b} developed a simple model
for the redistribution of energy and angular momentum in proto-binary
systems such that it directly leads to binary properties
(mass ratio, eccentricity and period) at the moment the star cluster
dynamical evolution becomes effective. In practice, this process
creates the IBP that should be used at the starting point of star cluster 
simulations. This process of converting birth binaries into initial binaries
is usually called {\it pre-main-sequence eigenevolution}, since
it occurs during the pre-main-sequence phase of binaries.

The first distinction we make is between the terms {\it birth} and
{\it initial} binary populations. The term {\it birth} here is applied, as 
in \citet{Kroupa_1995b}, to all protostars that are embedded in
circumprotostellar material. On the other hand, the term {\it initial}
corresponds to pre-main-sequence stars not embedded anymore in such
a material, it being accreted or expelled during the process of 
redistribution of energy and angular momentum. We use the
subscripts `ini' and `bir' to refer to the initial and birth population, 
respectively. 
The birth population can be viewed as a theoretical devise
to allow calculations, and it is not observable because proto-binaries
evolve rapidly on a time-scale of $10^5$ yr in a deeply embedded phase.
This is similar to the concept of initial mass function (i.e. the 
distribution of birth-stellar masses), which is not an observable distribution
function and also does not exist, but it is an important tool for computations
\citep{Kroupa_2013}. With such a tool, predictions can be made for observable
populations, which is why such theoretical distribution functions are needed.

The second distinction we consider is related to {\it low-mass} binaries 
and {\it high-mass} binaries. Low-mass binaries in this work correspond
to all binaries whose primary mass is smaller than $5$ M$_\odot$. Binaries
with primary masses greater than this are called high-mass binaries. Note
that we always assume that primaries are more massive than secondaries,
and that mass ratios are always smaller than or equal to unity 
(i.e. $q=M_2/M_1 \leq 1$). 

The third distinction concerns {\it short-period} binaries and {\it long-period}
binaries. Short-period binaries have $P < 10^3$ days while long-period binaries 
have $P \geq 10^3$ days.

Another distinction we make is related to the {\it original} Kroupa IBP and 
the {\it modified} Kroupa IBP. The original Kroupa IBP corresponds to
the IBP generated  through pre-main-sequence eigenevolution as described in \citet{Kroupa_1995b},
for low-mass binaries and to the IBP described in \citet{Kroupa_2013}, for high-mass
binaries. As we will see in Section \ref{problem}, an inconsistency appears
in simulated GC CMDs while comparing with real GC CMDs when the 
original Kroupa IBP is adopted. In order to overcome this 
problem, in this paper we propose a revised pre-main-sequence eigenevolution formulation, which leads in turn to 
a modified Kroupa IBP. Additionally, the modified Kroupa IBP, for high-mass 
binaries, assumes observed distributions for binaries whose primaries are O 
and B-dwarfs.

As a final concept, we use the term {\it stimulated evolution} to refer
to the process of cluster dynamical evolution such that it converts
the IBP to binaries whose distributions resemble those observed in
the Galactic field. Usually, stimulated evolution is relatively
fast and lasts for just a few Myr. In addition, residual gas removal due to
the evolution of the most massive stars leading to significant cluster expansion
\citep[e.g.][]{Brinkmann_2016} is assumed to stop stimulated evolution. 

As usually accepted, we assume that the Galactic field stellar population has its origin
in clustered star formation (i.e. dynamical processing of binaries
takes place before the dissolved cluster becomes part of the Galactic
field. In addition, we assume that such clusters contain a high binary
fraction \citep[e.g.][]{Duchene_1999,Kroupa_INITIAL}. 
Actually, we assume $\approx$ 100 
per cent of binaries, which 
is consistent with resolving the angular momentum problem in star formation
and the result that triples and higher order systems are rarely
the outcome of late-type star formation \citep{Goodwin_2005},
such that it leads to an appropriate binary fraction and binary properties
for different spectral type stars after stimulated evolution.



\section{ORIGINAL KROUPA IBP}
\label{original}

In this section we describe the features of the original Kroupa IBP.
These include properties of low-mass binaries (Section \ref{latetype1})
and high-mass binaries (Section \ref{earlytype1}). In the end, in Section 
\ref{problem}, we motivate the necessity of changing slightly the original 
prescriptions based on mock observations of present-day GC CMDs.


\subsection{Low-mass binaries (primary star mass < 5 M$_\odot$)}
\label{latetype1}

Note that the pre-main-sequence eigenevolution was developed
in order to explain observational correlations found for G, K and M-dwarf 
binaries in the Galactic field \citep{DM_1991,M_1992,FM_1992}, and was
ultimately confirmed when data of all late-type binary systems near 
the Sun became available  \citep{RG_1997}. Therefore,  
pre-main-sequence eigenevolution is a process associated closely with late-type or
low-mass binaries.

In what follows, we describe pre-main-sequence eigenevolution
in its original formulation, i.e. \citet{Kroupa_1995b}.


\subsubsection{Birth population}
\label{birth1}

In order to pass though pre-main-sequence eigenevolution, birth
binaries are born with specific distributions, assumed here as:

\begin{description}
\item[i)] {\it Primary mass}: randomly chosen from the Kroupa canonical initial mass function (IMF) \citep{Kroupa_1991}.
\item[ii)] {\it Secondary mass}: secondary is randomly chosen from the same IMF (i.e. the mass
ratio distribution is such that the binary components are randomly paired). 
Note that the primary and secondary
are evident only after both stars are chosen independently from the canonical IMF.

\item[iii)]{\it Eccentricity}: it follows a thermal distribution, i.e.

\begin{equation}
f_e \ = \ 2e \; . \label{EQ2}
\end{equation}

Here ${\rm d}N = f_e {\rm d}e$ is the fraction of orbits with
eccentricity in the range $e$ to $e+{\rm d}e$, amongst all orbits. 

\item[iv)] {\it Period}: it follows Eq. 8 in \citet{Kroupa_1995b}, i.e.

\begin{equation}
f_P \ = \ 2.5 \ \frac{\log_{10}(P/{\rm days}) - 1}{45 + \left[ \ \log_{10}(P/{\rm days}) \ - \ 1 \ \right]^2}  . \label{EQ3}
\end{equation}

Here ${\rm d}N = f_P {\rm d} \log_{10} P$ is the fraction of orbits with period in the range 
$\log_{10} P$ to $\log_{10} P + {\rm d}\log_{10} P$ such that the integral of $f_P$ over
all $\log_{10} P$ values equals the binary fraction of the stellar population, being one in this case.

\end{description}

During the pre-main-sequence eigenevolution phase, significant changes
in the binary properties occur when the less massive object (secondary) is
at the pericentre ($R_{p}$), i.e., the distributions above (i--iv) applied
for the birth population can be drastically changed during subsequent
passages through the pericentre. A convenient way to calibrate
the strength of the pre-main-sequence eigenevolution is by means of a function of $R_{p}$.
This is expressed in the function $\rho$, which is defined as

\begin{equation}
\rho \ = \ \left( \frac{\lambda \, {\rm R}_\odot}{R_{p}} \right)^{\chi}, \label{EQ4}
\end{equation}
where $\lambda = 28$ and $\chi = 3/4$ \citep{Kroupa_1995b}, and $R_{p}$ is in units of R$_\odot$. 
Note that the larger the pericentre distance, the smaller is the value of 
$\rho$.


\subsubsection{Initial population}
\label{initial1}

During the pre-main-sequence eigenevolution which has a duration of about $10^5$ yr, the birth population is converted to the
initial population. The change in the eccentricity due to pre-main-sequence eigenevolution 
is given by

\begin{equation}
\ln(e_{\rm ini}) \ = \ - \ \rho \ + \ \ln(e_{\rm bir}) . \label{EQ5}
\end{equation}

As during passages through the pericentre the secondary might accrete 
matter from the circumstellar disc around the primary, the changes in the
mass ratio due to the pre-main-sequence eigenevolution is

\begin{equation}
q_{\rm ini} = \left\{
\begin{array}{rcl}
q_{\rm bir} + \rho (1 - q_{\rm bir}),& \mbox{if} &  \rho \leq 1 ,\\
1                                   ,& \mbox{if} &  \rho > 1 . \\
\end{array}
\right. \label{EQ6}
\end{equation}

And the change of the secondary mass is given by 
\begin{equation}
M_{2,\rm ini} \ = \ q_{\rm ini} \, M_{1,\rm bir} .
\end{equation}

Note that the primary mass remains unchanged during pre-main-sequence eigenevolution, i.e.
$M_{1,\rm ini} = M_{1,\rm bir}$ since the secondary unlikely has an appreciable
circumstellar disk \citep{Bonnell_1992}.

Finally, the period after the pre-main-sequence eigenevolution is given by

\begin{equation}
P_{\rm ini} \ = \ P_{\rm bir} \ \left( \frac{M_{1,\rm bir} + M_{2,\rm bir}}{M_{1,\rm ini} + M_{2,\rm ini}} \right)^{1/2} \ \left( \frac{1 - e_{\rm bir}}{1 - e_{\rm ini}} \right)^{3/2} . \label{EQ7}
\end{equation}

We must comment a few things at this point before proceeding further.

First, concerning mergers during the process of pre-main-sequence eigenevolution, binaries
with pericentre distances greater than $1.1 \times (R_1 + R_2)$ survive, and
they merge otherwise, where $R_1$ and $R_2$ are the primary and secondary radii,
respectively. 

Second, the pre-main-sequence eigenevolution changes mainly short-period binaries
since for these systems the function $\rho$ can be truly large.
For long-period binaries, the effect of pre-main-sequence eigenevolution is inexpressive because
the $\rho$ is extremely small (due to large pericentre distance) which leads to 
initial properties similar to birth properties.

Third, given the large amount of late-type pre-main-sequence objects generated 
according to the adopted IMF, many short-period low-mass binaries might have equal 
masses and low eccentricity due to pre-main-sequence eigenevolution, since they more easily have 
their properties changed. This has a huge impact in a star cluster population 
evolved over a Hubble time (Section \ref{problem}) as these binaries are likely 
to survive for several Gyr of cluster dynamical evolution. 

\begin{figure}
   \begin{center}
    \includegraphics[width=0.99\linewidth]{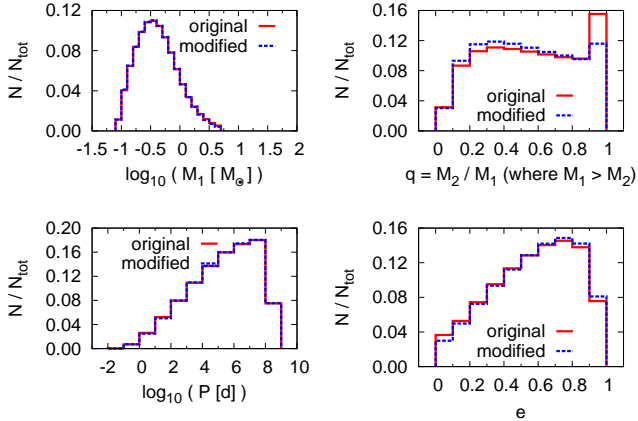} 
    \end{center}
  \caption{Primary mass (top left-hand panel), mass ratio (top right-hand panel), 
period (bottom left-hand panel), and eccentricity (bottom right-hand panel) distributions
for all binaries such that $M_1 < 5$ M$_\odot$ in the original Kroupa IBP (Section 
\ref{latetype1}, red solid line) and in the modified Kroupa IBP (Section \ref{latetype2},
blue dashed line).}
  \label{FIG01}
\end{figure}

In Fig. \ref{FIG01} we plot the main distributions associated with the original Kroupa 
IBP regarding low-mass binaries, i.e. primary mass (top left-hand panel), mass ratio 
(top right-hand panel), period  (bottom left-hand panel), and eccentricity (bottom 
right-hand panel). Note that we included as well the modified Kroupa IBP (Section 
\ref{latetype2}) for comparison.


\begin{figure}
   \begin{center}
    \includegraphics[width=0.99\linewidth]{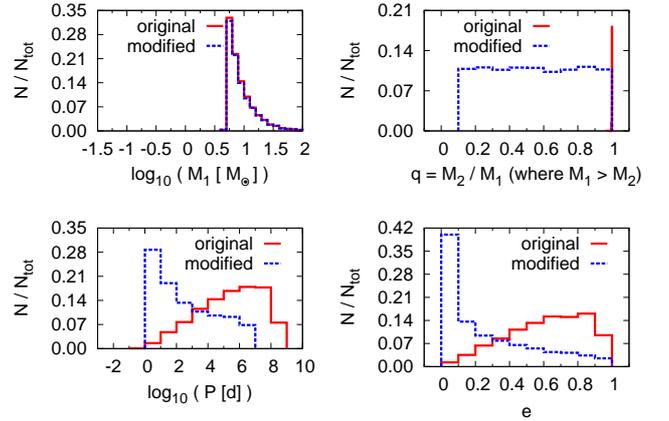} 
    \end{center}
  \caption{Primary mass (top left-hand panel), mass ratio (top right-hand panel), 
period (bottom left-hand panel), and eccentricity (bottom right-hand panel) distributions
for all binaries such that $M_1 > 5$ M$_\odot$ in the original Kroupa IBP (Section 
\ref{earlytype1}, red solid line) and in the modified Kroupa IBP (Section \ref{earlytype2},
blue dashed line).}
  \label{FIG02}
\end{figure}

\subsection{High-mass binaries (primary star mass > 5 M$_\odot$)}
\label{earlytype1}

Since the pre-main-sequence eigenevolution theory was developed mainly to explain
observed properties of Galactic field low-mass binaries (A, F, G, K and M-dwarfs),
the question concerning how high-mass binaries (O and B-dwarfs)\footnote{Based 
on our definition for low-mass and high-mass binaries, binaries whose
primaries are A-dwarfs are low-mass binaries, even though they are early-type binaries. 
We follow here the same threshold as in \citet{Kroupa_2013}.} evolve remains open.

According to \citet{Kroupa_2013}, the initial properties of high-mass binaries follow
precisely birth distributions, since pre-main-sequence eigenevolution is not applied in this
range of masses. 
This is a reasonable assumption since for this range of masses, the timescale
of pre-main-sequence evolution is short enough that it is safe to neglect it
\citep{Railton_2014}.
Then, for high-mass binaries, birth and initial binaries
follow distributions (i--iv) in Section \ref{latetype1}, with one exception,
namely the secondary mass (i.e. mass ratio). Until very recently, observations
of high-mass binaries have commonly revealed systems with nearly equal masses
\citep[e.g.][]{Pinsonneault_2006,Mayer_2017}. 

In order to account for this feature, in the original Kroupa IBP, a different
pairing is usually adopted for high-mass binaries, namely {\it ordered} pairing
\citep{Kroupa_2013}. 
The procedure can be summarized as follows. Once all stars (twice the 
number of binaries) are generated according to the canonical Kroupa IMF, those with
masses greater than $5$ M$_\odot$ are subsequently ordered. The most massive star in 
the array is paired with the second most massive stars, and so on. This procedure 
guarantees that the IMF is preserved and generates high-mass binaries with similar masses, 
as previously thought from observational results. 

Fig. \ref{FIG02} exhibits the main distributions associated with the original Kroupa
IBP with respect to high-mass binaries, i.e. primary mass (top left-hand panel), mass 
ratio (top right-hand panel), period  (bottom left-hand panel), and eccentricity 
(bottom right-hand panel). Note that we included as well the modified Kroupa IBP 
(Section \ref{earlytype2}) for comparison, as in Fig. \ref{FIG01}.


\section{NUMERICAL TOOLS}
\label{codes}

We describe is this section the two codes used in this investigation
for star cluster simulations ({\sc mocca}, Section \ref{mocca}) and
for the photometric reduction of mock observations of the simulated star clusters ({\sc cocoa},
Section \ref{cocoa}).


\subsection{The {\sc mocca} code}
\label{mocca}

The {\sc mocca} code \citep[][and references therein]{Hypki_2013,Giersz_2013}
is based on the orbit-averaged Monte Carlo technique
 for cluster  evolution developed by \citet{Henon_1971}, which 
was further improved by  \citet{Stodolkiewicz_1986}. It also includes the 
{\sc fewbody} code, developed by \citet{Fregeau_2004}, to perform 
numerical scattering experiments of gravitational interactions.
To model the Galactic potential, {\sc mocca} assumes a point-mass with 
total mass equal to the enclosed Galaxy mass at the Galactocentric 
radius. The description of the escape processes in tidally limited 
clusters follows the procedure derived by 
\citet{Fukushige_2000}. Stellar evolution is implemented via
the {\sc sse} code developed by \citet{Hurley_2000} for single stars and
the {\sc bse} code developed by \citet{Hurley_2002} for binary evolution. 

{\sc mocca} was extensively tested against $N$-body codes. 
For instance, \citet{Giersz_2013} concluded that {\sc mocca} is 
capable of reproducing $N$-body results with reasonable precision, not 
only for the rate of cluster evolution and the cluster mass distribution, 
but also for the detailed distributions of mass and binding energy of 
binaries. Additionally, \citet{Wang_2016} also compared {\sc mocca} with
the state-of-the-art {\sc nbody6++gpu} and showed good agreement
between the two codes. Finally, Madrid et al. (in prep.) showed that
the {\sc mocca} code is able to reproduce the escape rate from tidally
limited clusters when compared with $N$-body codes, provided that
they are not closer than a few kpc from the Galactic center.


\subsection{The {\sc cocoa} code}
\label{cocoa}

The {\sc cocoa} code \citep[][]{Askar_2017b} can create idealised mock observational data
in the form of FITS files using numerical simulation snapshots at a specific time
(provided by codes such as {\sc mocca}/{\sc nbody6}/{\sc nbody6++gpu}). The {\sc cocoa} code 
has been developed to extend results of numerical simulations of star clusters for the purpose of 
direct comparisons with observations. The input parameters in {\sc cocoa} can be adjusted to create 
synthetic observations from virtually any optical telescope and the code can also carry out PSF photometry on 
the mock observations to create a catalogue of all observed stars in the cluster. These results can be used to 
observationally determine cluster parameters and create observational CMDs of a simulated star cluster model.

The {\sc cocoa} code has already been used for creating mock observations using 12 Gyr simulation snapshots
in investigations with the {\sc nbody6++gpu} code \citep{Wang_2016} and the {\sc mocca} code \citep{Askar_2017a}.
{\sc cocoa} has many different applications \citep{Askar_2017b} and can check if there are any systematics or 
biases associated with actual observational data and techniques used to determine cluster properties.


\section{PROBLEM WITH THE ORIGINAL KROUPA IBP}
\label{problem}

The original Kroupa IBP (Section \ref{original})
has been tested against both numerical simulations and observations,
and has successfully explained the observational features of young clusters,
associations, Galactic field late-type binaries, and even binaries in old GCs
\citep[e.g.][and references therein]{Kroupa_2011,Marks_2012,Leigh_2015}.

Even though the original prescription gives good results, there is
at least one problem with the way pre-main-sequence eigenevolution changes the birth population.
As noted in Section \ref{latetype1}, the pre-main-sequence eigenevolution might be very
strong for short-period G, K and M-dwarfs, in the sense that their 
initial properties can be extremely different from their birth properties. 
For example, substantial number of these short-period binaries will have equal 
masses and very low-eccentricities. This makes them very dynamically hard 
binaries which in turn allows them to survive in star clusters during 
long-term dynamical evolution (see Fig. \ref{FIG03.3}).

The main implication of the above-mentioned result of the pre-main-sequence eigenevolution 
in low-mass short-period binaries in GCs are MS binary mass ratio distributions characterized
by a significant fraction of binaries with $q \approx 1$, 
which are not typical in observed GCs \citep{Milone_2012}.

First, in Section \ref{problem_model}, we evolve 6 GC models (which follow
initially the Kroupa IBP) with the {\sc mocca} 
code and perform the photometric analysis in those models with the {\sc cocoa} code.
Then we show that the colour distributions in the synthetic CMDs, for all
models, have conspicuous peaks towards the right edge of the distributions.
Such peaks are associated with binaries
whose mass ratios are $\approx$ 1, as discussed above.

Second, in Section \ref{problem_real}, we show CMDs of two real GCs and also
their colour distributions, following the same procedure employed in the
synthetic CMDs. We end this section by showing that the Kroupa IBP predicts 
colour distributions (and in turn MS binary mass ratio distributions) that 
are unlikely to exist in real GCs.


\subsection{GC models, synthetic CMDs and colour distributions}
\label{problem_model}

\begin{figure}
   \begin{center}
    \includegraphics[width=0.48\linewidth]{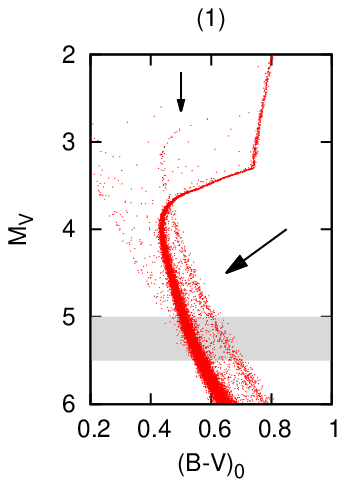}
    \includegraphics[width=0.48\linewidth]{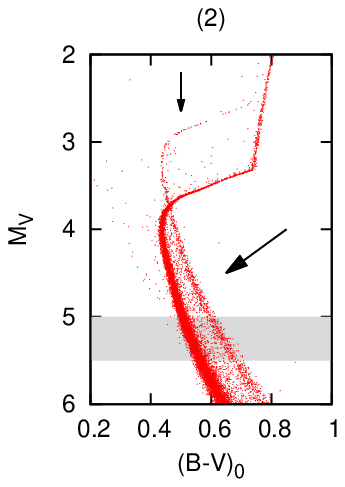}
    \includegraphics[width=0.48\linewidth]{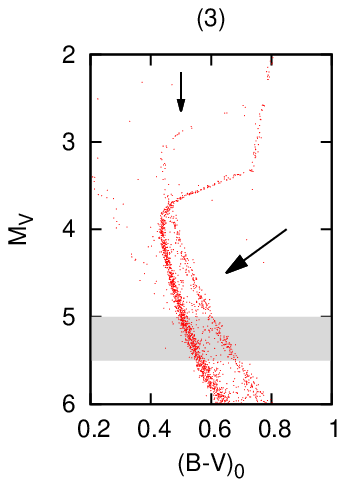}
    \includegraphics[width=0.48\linewidth]{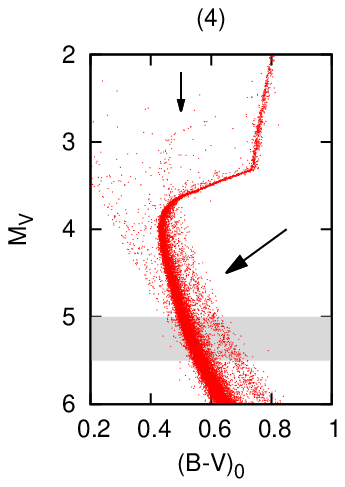}
    \includegraphics[width=0.48\linewidth]{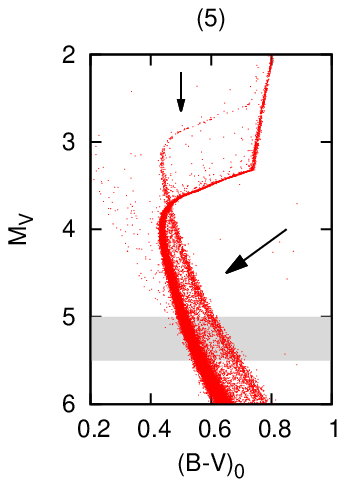}
    \includegraphics[width=0.48\linewidth]{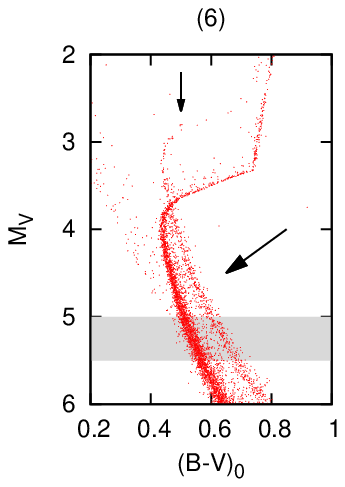}
    \end{center}
  \caption{Present-day CMDs from simulated observations of the six GC models 
(Table 1) evolved with the original Kroupa IBP (Section \ref{original}). 
Regions used to generate colour distributions in Fig. \ref{FIG03.2} are demarcated by gray areas.
Note that there is a pronounced binary sequence (particularly above the MS turn-off ) due 
to short-period low-mass binaries with mass ratios of unity in the model that has been 
indicated by the black arrows in the figure. 
For more detail see Section \ref{problem_model}.
}
  \label{FIG03.1}
\end{figure}

\begin{figure}
   \begin{center}
    \includegraphics[width=0.49\linewidth]{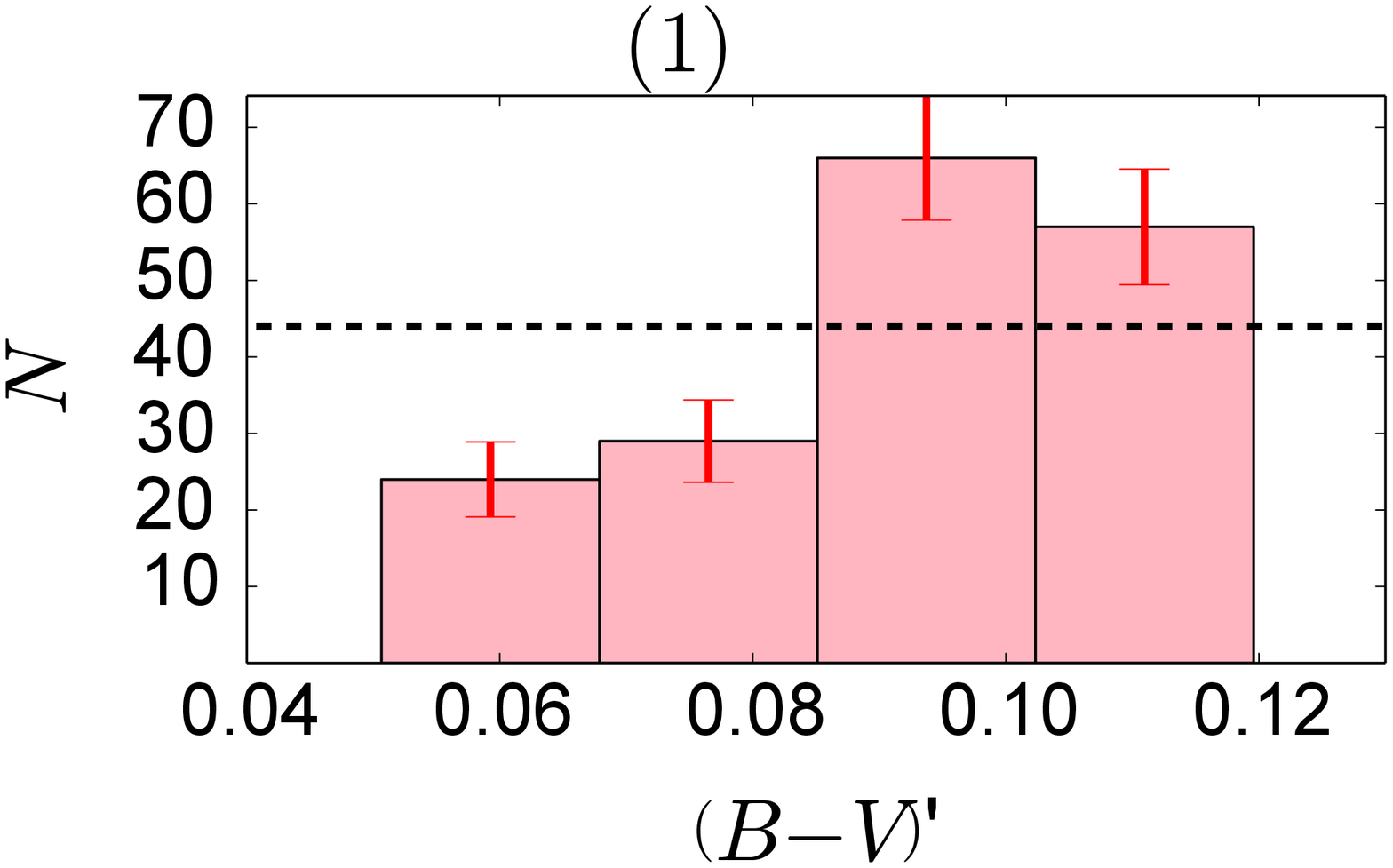} 
    \includegraphics[width=0.49\linewidth]{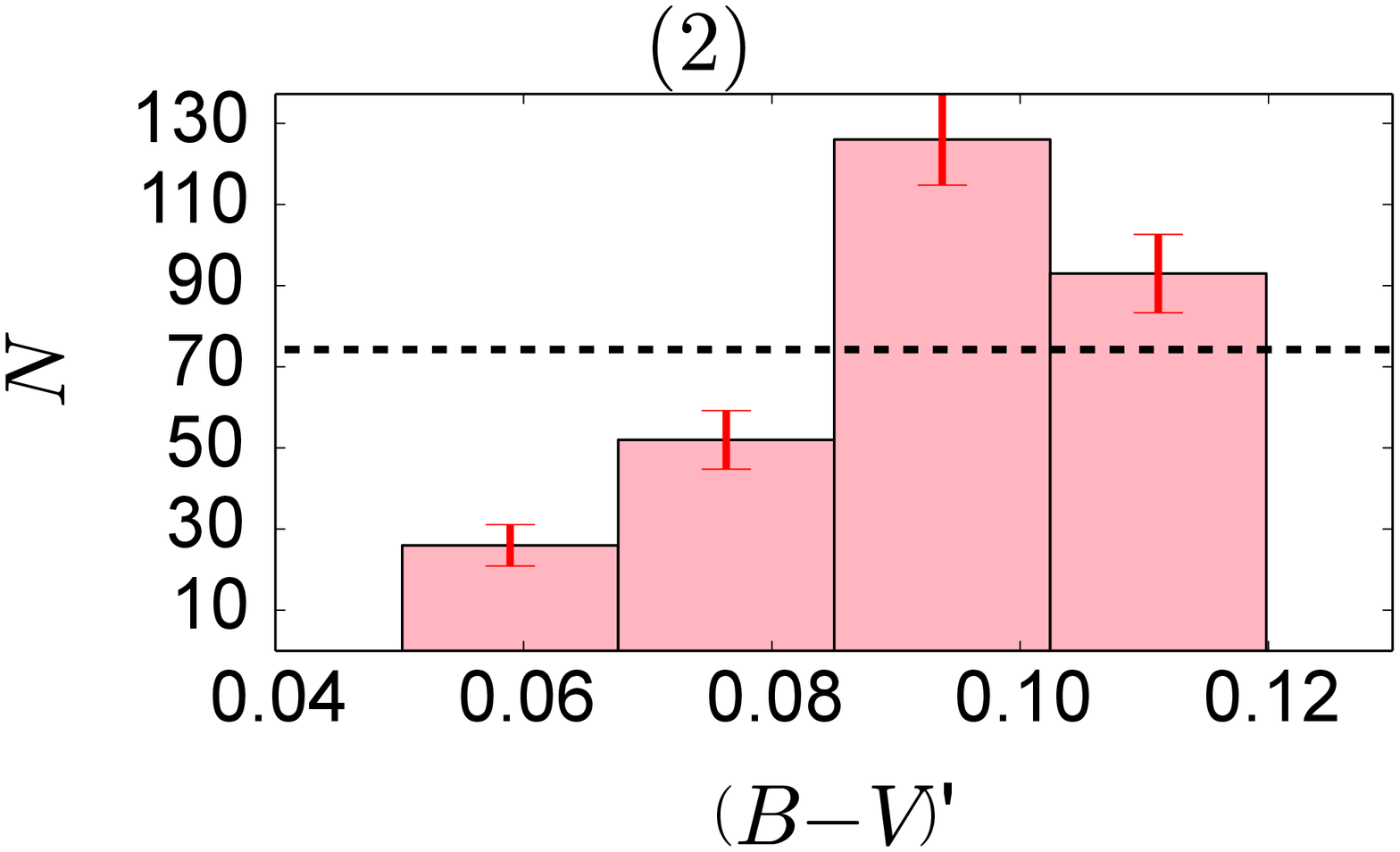} 
    \includegraphics[width=0.49\linewidth]{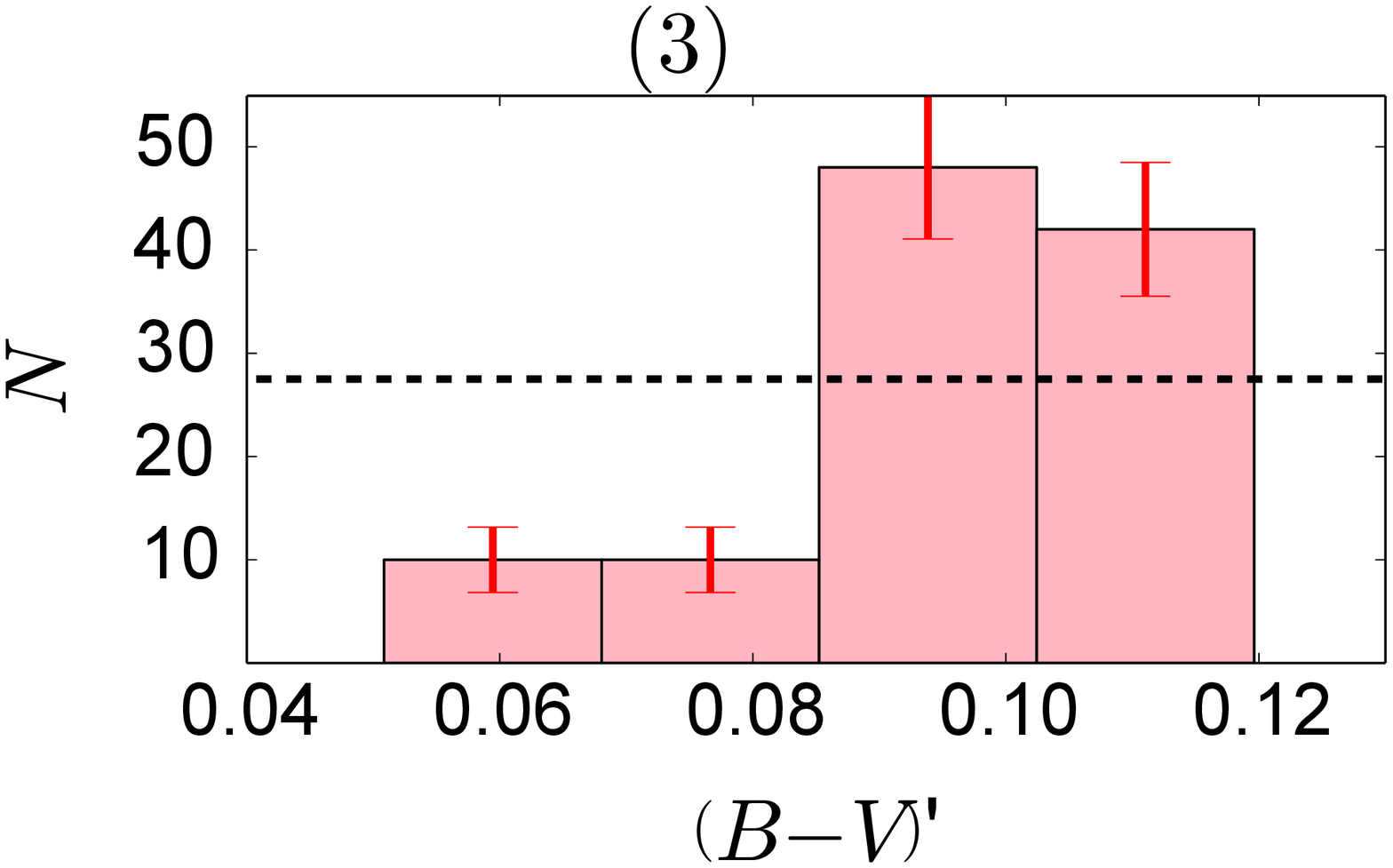} 
    \includegraphics[width=0.49\linewidth]{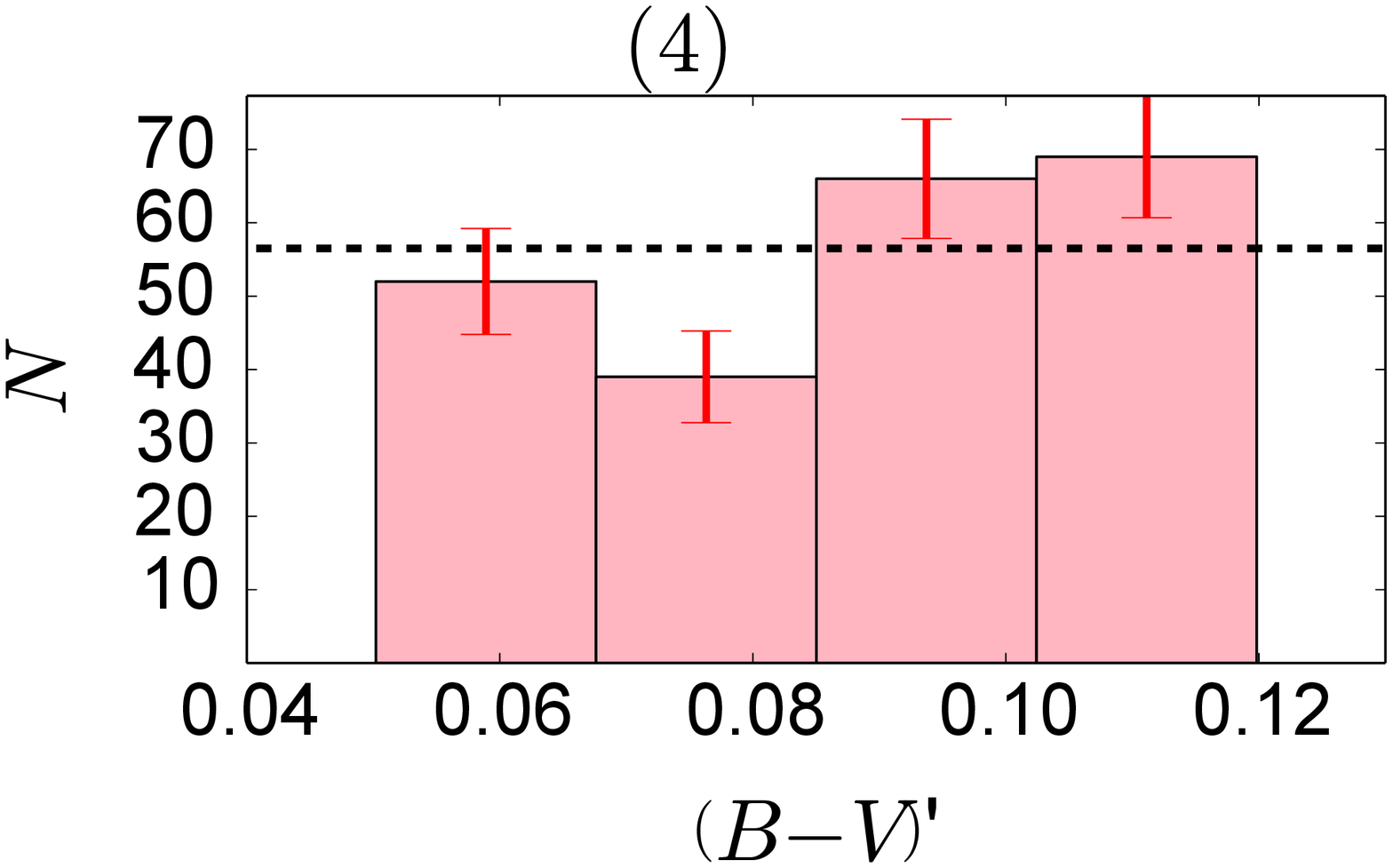} 
    \includegraphics[width=0.49\linewidth]{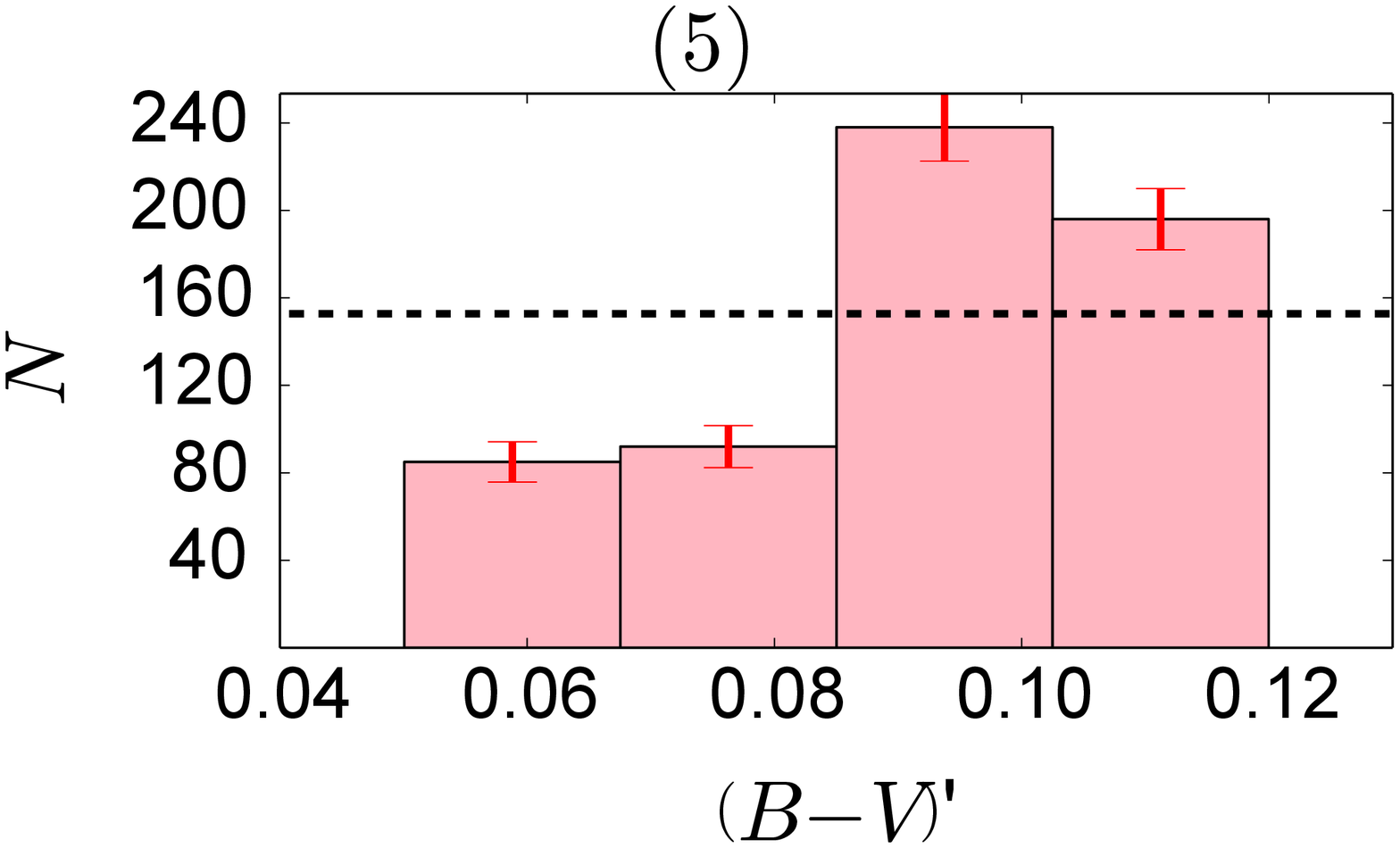} 
    \includegraphics[width=0.49\linewidth]{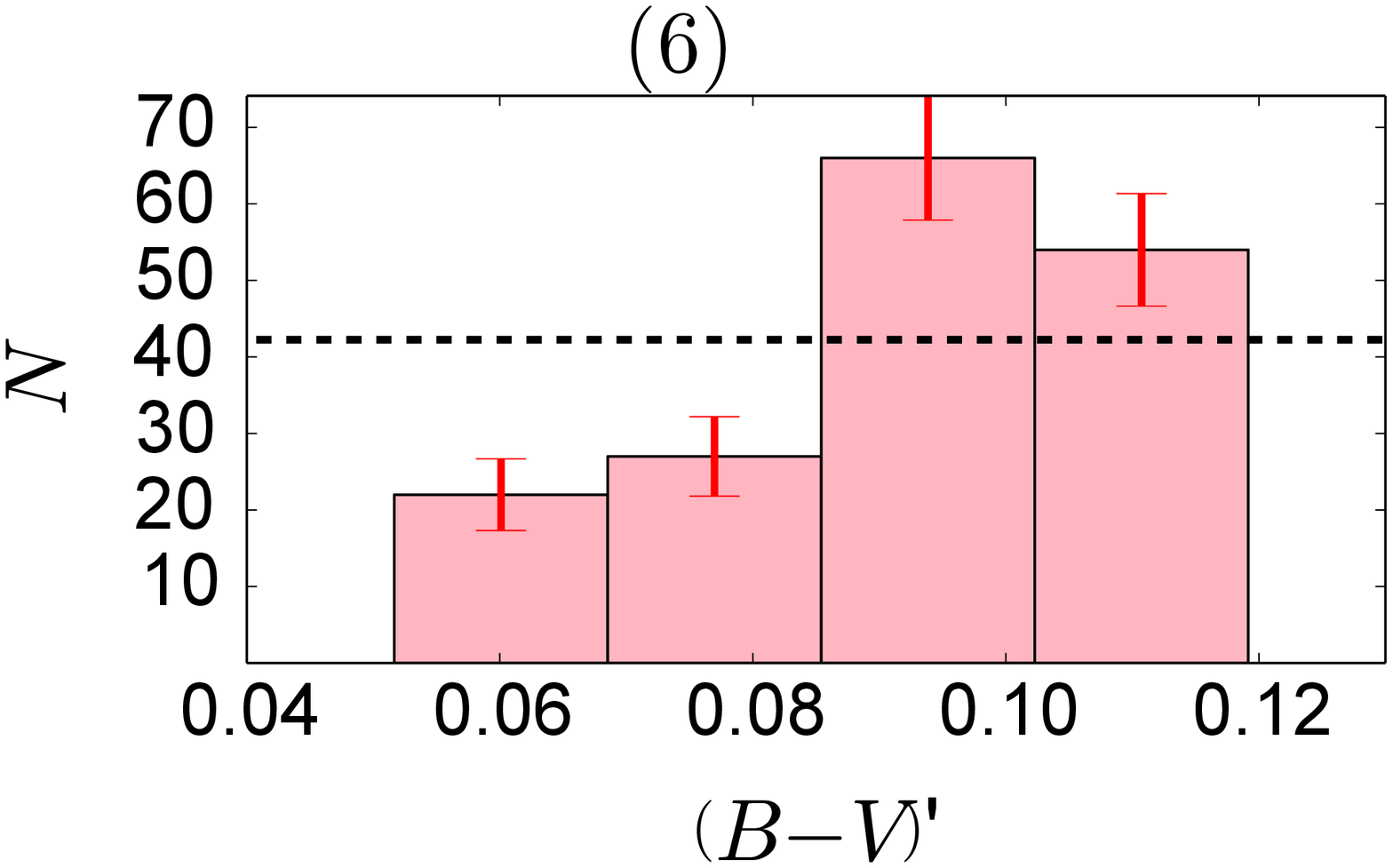} 
    \includegraphics[width=0.98\linewidth]{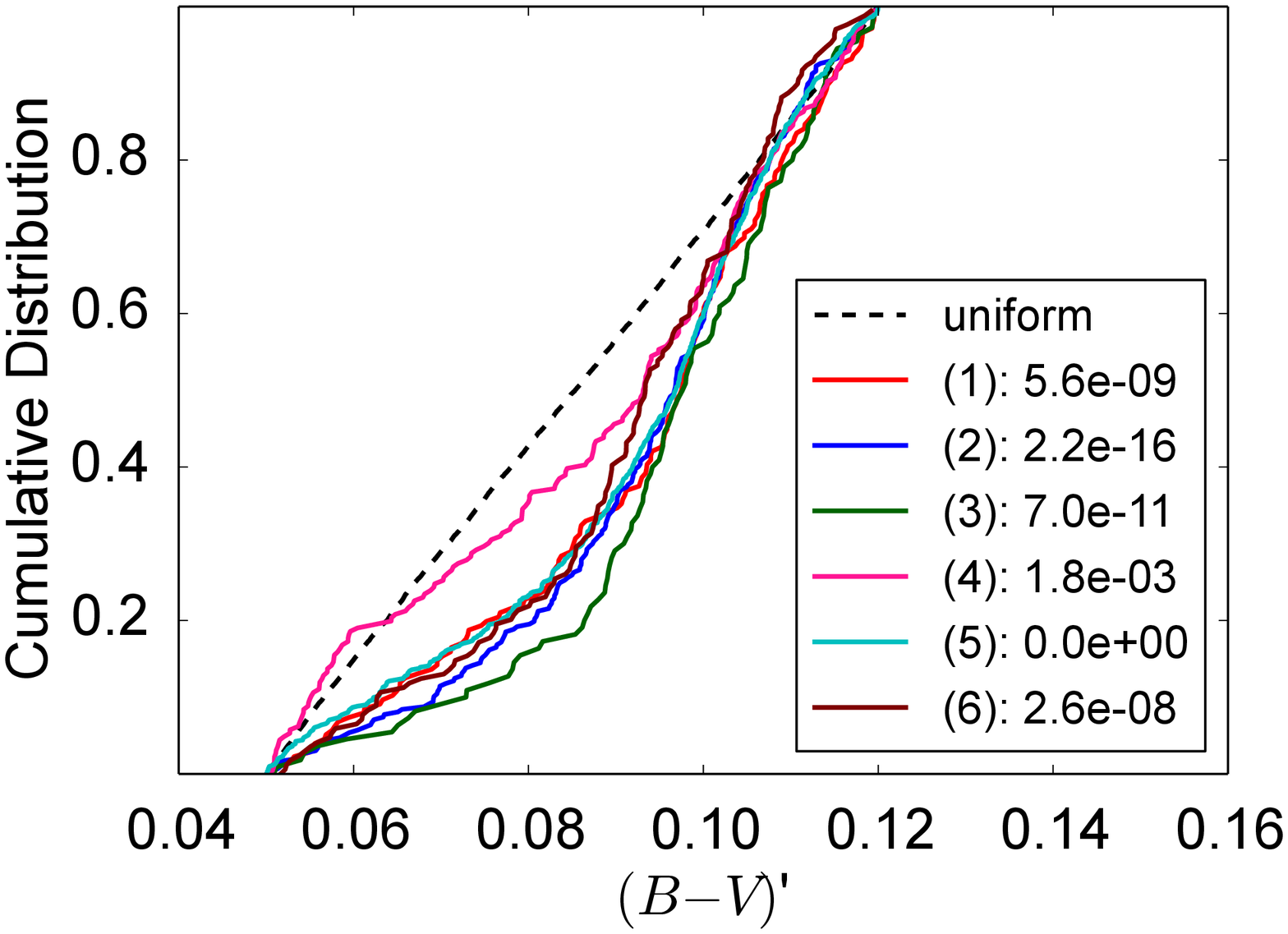}
    \end{center}
  \caption{Colour distributions derived from the CMDs shown in Fig. \ref{FIG03.1},
following the procedure described in Section \ref{problem_model}.
In the first six panels, we show histograms together with horizontal lines which are 
average numbers, assuming flat  distributions. Vertical lines in the histograms correspond 
to Poisson errors. In the last panel,
we compare cumulative distributions, and display in the key $p$-values of 
one-sample Kolmogorov-Smirnov tests for uniformity, for each model.
Only stars with $(B-V)'>0.05$ are used to compute the above distributions.
Note that in all colour distributions a conspicuous peak is observed towards 
the right edge the distributions, which is associated with the binary sequences visible in the CMDs
exhibited in Fig. \ref{FIG03.1}.
With respect to the cumulative distributions,
it is clear that the colour distributions in the six models following the original
Kroupa IBP are not uniform. This is also supported by the statistical test, which allows us to
reject the hypothesis that they are uniform with more than 99 \% of confidence (see $p$-values). 
For more detail see Sections \ref{problem_model}.
}
  \label{FIG03.2}
\end{figure}

\begin{figure}
   \begin{center}
    \includegraphics[width=0.9\linewidth]{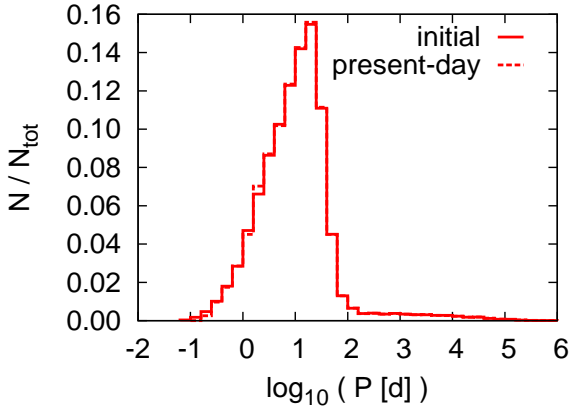}  
    \end{center}
  \caption{Initial and present-day period distributions of all MS binaries 
whose $q>0.99$, in all six models described in Table 1. 
Note that both distributions are practically the same, which strongly supports
the idea that the binary sequence for binaries with $q \approx 1$ comes from
primordial binaries practically not affected by dynamics. The reason is because
such binaries are very-short-period, having the majority of binaries with 
$q \approx 1$ periods shorter than 100 days. This suggests that in order
to avoid the prediction  of such binary sequences, the IPB has to be changed. 
For more detail see Sections \ref{problem_model}.
}
  \label{FIG03.3}
\end{figure}

\begin{sidewaystable*}
\centering
\begin{adjustbox}{max width=\textwidth}
\label{Tab01}
\noindent
\begin{threeparttable}
\noindent
\setlength\tabcolsep{1.5pt} 
\begin{tabular}{l|c|c|c|c|c|c|c|c|c|c|c|c|c|c|c|c|c|c|c|c|c|c|c}
\caption{GC models, initial parameters that define them and present-day properties (i.e. after 12 Gyr of evolution). 
The binary fraction, the metallicity and the the King model parameter are 
95 per cent, 0.001 and 6.0, respectively, for all models,
In addition, for all models, we adopt the canonical \citet{Kroupa_1991}. IMF (i.e. with two segments),
and the original Kroupa IBP (Section \ref{original}).
The first column presents the model name and the following 9
columns give the initial values for the total
mass (in M$_\odot$), the total number of stars, the central density (in M$_\odot$ pc$^{-3}$), 
the core radius (in pc), the half-mass radius (in pc),
the tidal radius (in pc), the Galactocentric
distance (in kpc), the half-mass radius relaxation time (in Myr),
and the central velocity dispersion (in km s$^{-1}$). The remaining 11 columns provide
present-day values for the total
mass (in M$_\odot$), the total number of stars, the central density (in M$_\odot$ pc$^{-3}$), 
the core radius (in pc), the half-mass radius (in pc),
the tidal radius (in pc), the half-mass radius relaxation time (in Myr),
the non-mass-weighted central velocity dispersion (in km s$^{-1}$), the cluster type \tnote{a} and 
the total cluster binary fraction.
See Section \ref{mocca}, for more details.} \label{Tab01} \\

\hline
  & \multicolumn{10}{c}{initial} & & \multicolumn{9}{c}{present-day}  \\
\hline
Model &
\specialcell{M\\$\left[  \times \, 10^5 \, {\rm M}_\odot \right]$} & 
\specialcell{N\\$\left[ \times \, 10^5 \right]$} & 
\specialcell{$\rho_c$\\$\left[ {\rm M}_\odot \; {\rm pc}^{-3} \right]$} &   
\specialcell{$R_{\rm c}$\\$[{\rm pc}]$}  & 
\specialcell{$R_{\rm h}$\\$[{\rm pc}]$} &  
\specialcell{$R_{\rm t}$\\$[{\rm pc}]$}  & 
\specialcell{$R_{\rm G}$\\$[{\rm kpc}]$}  & 
\specialcell{$T_{\rm r,h}$\\$\left[ {\rm Myr} \right]$} & 
\specialcell{$\sigma_c$\\${\rm \left[ km \, s^{-1} \right]}$} &
 &
\specialcell{M\\$\left[  \times \, 10^5 \, {\rm M}_\odot \right]$} & 
\specialcell{N\\$\left[ \times \, 10^5 \right]$} & 
\specialcell{$\rho_c$\\$\left[ {\rm M}_\odot \; {\rm pc}^{-3} \right]$} &   
\specialcell{$R_{\rm c}$\\$[{\rm pc}]$}  & 
\specialcell{$R_{\rm h}$\\$[{\rm pc}]$} &  
\specialcell{$R_{\rm t}$\\$[{\rm pc}]$}  & 
\specialcell{$T_{\rm r,h}$\\$\left[ {\rm Myr} \right]$} & 
\specialcell{$\sigma_c$\\${\rm \left[ km \, s^{-1} \right]}$} & 
Type \tnote{a} &
\specialcell{Total binary\\fraction} \\

\hline\hline

1 &  3.70 &  7.8 &  2.43 $\times 10^4$  & 0.71 &  2.4 &60&  4.4 & 5.60 $\times 10^2$ & 25.2 & 
& 2.07 &  6.2 & 3.6 $\times 10^7$  & 0.38 & 5.6 &  49.4 & 3.5 $\times 10^3$ &  34.5 & cIMBH & 0.12 \\ \hline

2 &  3.70 &  7.8 &  4.87 $\times 10^2$  & 2.62 &  8.8 &60&  4.4 & 3.95 $\times 10^3$ & 13.2 & 
& 1.06 &  3.0 & 1.9 $\times 10^2$  & 1.81 & 8.1 &  39.2 & 4.2 $\times 10^3$ &  6.1 &    pc & 0.24 \\ \hline

3 &  3.70 &  7.8 &  3.89 $\times 10^3$  & 1.31 &  4.4 &30&  1.6 & 1.40 $\times 10^3$ & 18.6 & 
& 0.09 &  0.2 & 2.4 $\times 10^5$  & 0.02 & 1.4 &   8.9 & 7.2 $\times 10^1$ &  4.5 &     c & 0.30 \\ \hline

4 &  9.25 & 19.5 &  1.05 $\times 10^5$  & 0.74 &  2.4 &60&  2.8 & 7.99 $\times 10^2$ & 30.7 & 
& 5.55 & 16.9 & 5.8 $\times 10^8$  & 0.58 & 4.9 &  50.6 & 4.6 $\times 10^3$ & 45.7 & cIMBH & 0.09 \\ \hline

5 &  9.25 & 19.5 &  2.11 $\times 10^3$  & 2.73 &  8.8 &60&  2.8 & 5.64 $\times 10^3$ & 16.0 & 
& 2.91 &  8.5 & 3.6 $\times 10^2$  & 2.53 & 8.6 &  40.8 & 7.1 $\times 10^3$ & 10.1 &    pc & 0.20 \\ \hline

6 &  9.25 & 19.5 &  1.69 $\times 10^4$  & 1.37 &  4.4 &30&  1.0 & 1.99 $\times 10^3$ & 22.7 & 
& 1.15 &  2.5 & 3.3 $\times 10^4$  & 0.19 & 2.4 &  14.9 & 5.8 $\times 10^2$ &  11.0 &    pc & 0.16 \\ \hline

\hline
\label{Tab01}
\end {tabular}
\label{Tab01}
\begin{tablenotes}
       \item[a] The cluster present-day type can be: post-core collapse (c), post-core collapse with intermediate-mass black hole (cIMBH) and pre-core collapse (pc).
\end{tablenotes}
\label{Tab01}
\end{threeparttable}
\label{Tab01}
\end{adjustbox}

\qquad  
\vspace{2.0cm}

\begin{adjustbox}{max width=\textwidth}
\noindent
\begin{threeparttable}
\noindent
\setlength\tabcolsep{1.5pt} 
\begin{tabular}{l|c|c|c|c|c|c|c|c|c|c|c|c|c|c|c|c|c|c|c|c|c|c|c}
\label{Tab02}
\caption{GC models (whose initial conditions are exactly the same as those listed in Table 1), 
having initial binaries following the modified Kroupa IBP
(Section \ref{modified}). Columns are the same as in Table 1.
Notice that model 4, which contains a very massive IMBH (which dominates the dynamics of the core), 
has a significantly larger central velocity dispersion, when compared with the same model in Table 1. 
This is because the process of IMBH formation is stochastic \citep{Giersz_2015}, so the 
masses of IMBHs at 12 Gyr are different, which causes differences in non-mass-weighted central velocity dispersions.
See Section \ref{mocca}, for more details.}\\

\hline
  & \multicolumn{10}{c}{initial} & & \multicolumn{9}{c}{present-day}  \\
\hline
Model & 
\specialcell{M\\$\left[  \times \, 10^5 \, {\rm M}_\odot \right]$} & 
\specialcell{N\\$\left[ \times \, 10^5 \right]$} & 
\specialcell{$\rho_c$\\$\left[ {\rm M}_\odot \; {\rm pc}^{-3} \right]$} &   
\specialcell{$R_{\rm c}$\\$[{\rm pc}]$}  & 
\specialcell{$R_{\rm h}$\\$[{\rm pc}]$} &  
\specialcell{$R_{\rm t}$\\$[{\rm pc}]$}  & 
\specialcell{$R_{\rm G}$\\$[{\rm kpc}]$}  & 
\specialcell{$T_{\rm r,h}$\\$\left[ {\rm Myr} \right]$} & 
\specialcell{$\sigma_c$\\${\rm \left[ km \, s^{-1} \right]}$} &
 &
\specialcell{M\\$\left[  \times \, 10^5 \, {\rm M}_\odot \right]$} & 
\specialcell{N\\$\left[ \times \, 10^5 \right]$} & 
\specialcell{$\rho_c$\\$\left[ {\rm M}_\odot \; {\rm pc}^{-3} \right]$} &   
\specialcell{$R_{\rm c}$\\$[{\rm pc}]$}  & 
\specialcell{$R_{\rm h}$\\$[{\rm pc}]$} &  
\specialcell{$R_{\rm t}$\\$[{\rm pc}]$}  & 
\specialcell{$T_{\rm r,h}$\\$\left[ {\rm Myr} \right]$} & 
\specialcell{$\sigma_c$\\${\rm \left[ km \, s^{-1} \right]}$} & 
Type \tnote{a} &
\specialcell{Total binary\\fraction}\\

\hline\hline

1 &  3.70 &  7.8 &  2.43 $\times 10^4$  & 0.71 &  2.4 &60&  4.4 & 5.60 $\times 10^2$ & 25.2 & 
& 2.05 &  6.24 &  5.4 $\times 10^7$  & 0.21 & 5.5 &  49.4 & 3.5 $\times 10^3$ &  33.8 & cIMBH & 0.12 \\ \hline

2 &  3.70 &  7.8 &  4.87 $\times 10^2$  & 2.62 &  8.8 &60&  4.4 & 3.95 $\times 10^3$ & 13.2 & 
& 1.04 &  2.99 &  6.8 $\times 10^2$  & 3.62 & 8.1 &  39.5 & 4.2 $\times 10^3$ &  6.0 &    pc & 0.23 \\ \hline

3 &  3.70 &  7.8 &  3.89 $\times 10^3$  & 1.31 &  4.4 &30&  1.6 & 1.40 $\times 10^3$ & 18.6 & 
& 0.09 &  0.16 &  4.3 $\times 10^5$  & 0.05 & 1.3 &   8.9 & 6.9 $\times 10^1$ &  4.7 &     c & 0.26 \\ \hline

4 &  9.25 & 19.5 &  1.05 $\times 10^5$  & 0.74 &  2.4 &60&  2.8 & 7.99 $\times 10^2$ & 30.7 & 
& 5.48 & 16.98 &  8.6 $\times 10^8$  & 0.36 & 4.9 &  50.6 & 4.6 $\times 10^3$ &  66.9 & cIMBH & 0.08 \\ \hline

5 &  9.25 & 19.5 &  2.11 $\times 10^3$  & 2.73 &  8.8 &60&  2.8 & 5.64 $\times 10^3$ & 16.0 & 
& 2.76 &  8.22 &  2.5 $\times 10^3$  & 3.28 & 8.5 &  40.3 & 7.0 $\times 10^3$ &  9.6 &    pc & 0.20 \\ \hline

6 &  9.25 & 19.5 &  1.69 $\times 10^4$  & 1.37 &  4.4 &30&  1.0 & 1.99 $\times 10^3$ & 22.7 & 
& 1.04 &  2.25 &  2.5 $\times 10^4$  & 0.32 & 2.3 &  14.5 & 5.3 $\times 10^2$ & 10.8 &    pc & 0.16 \\ \hline

\hline
\end {tabular}
\begin{tablenotes}
       \item[a] The cluster present-day type can be: post-core collapse (c), post-core collapse with intermediate-mass black hole (cIMBH) and pre-core collapse (pc).
\end{tablenotes}

\end{threeparttable}
\end{adjustbox}

\end{sidewaystable*}

\begin{figure}
   \begin{center}
    \includegraphics[width=0.49\linewidth]{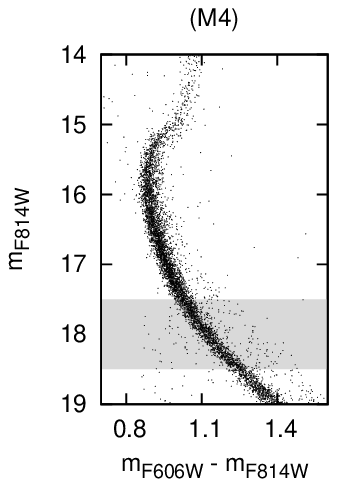} 
    \includegraphics[width=0.49\linewidth]{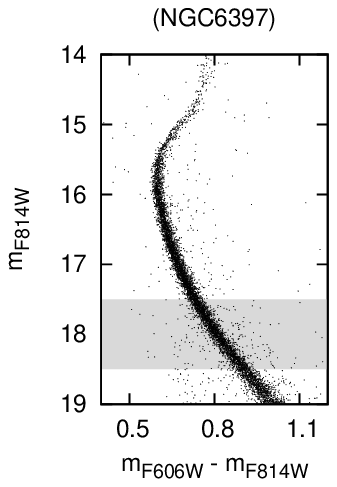} 
    \includegraphics[width=0.95\linewidth]{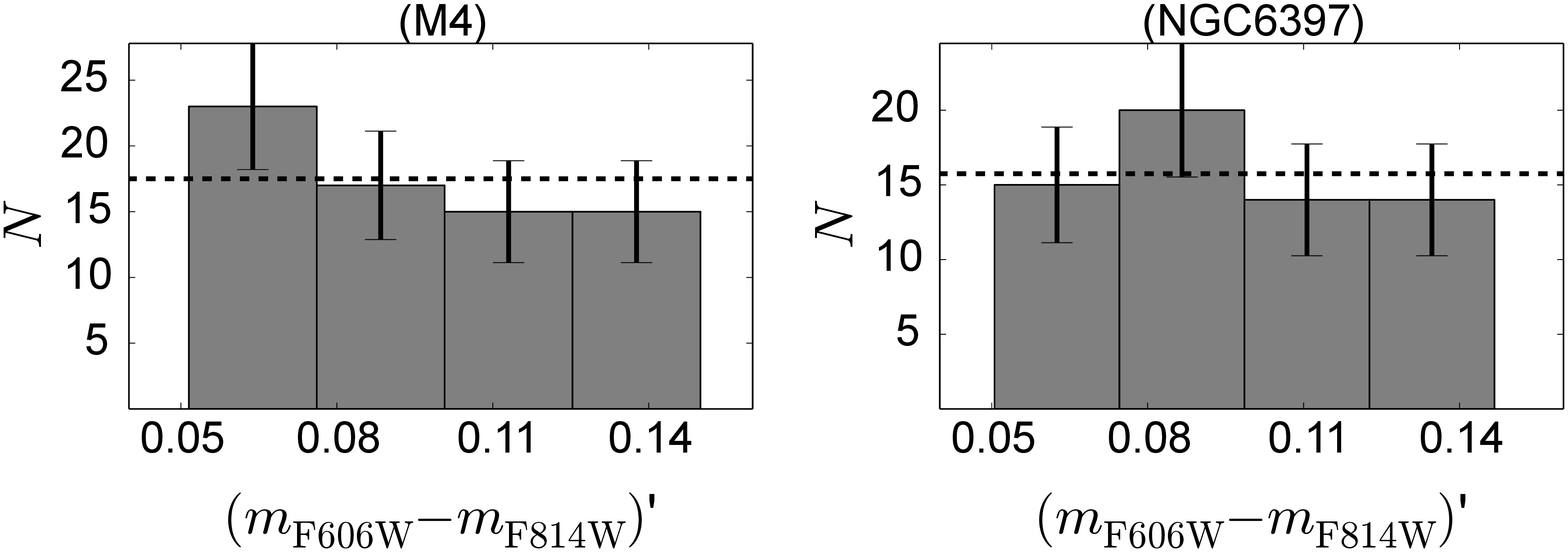} 
    \includegraphics[width=0.95\linewidth]{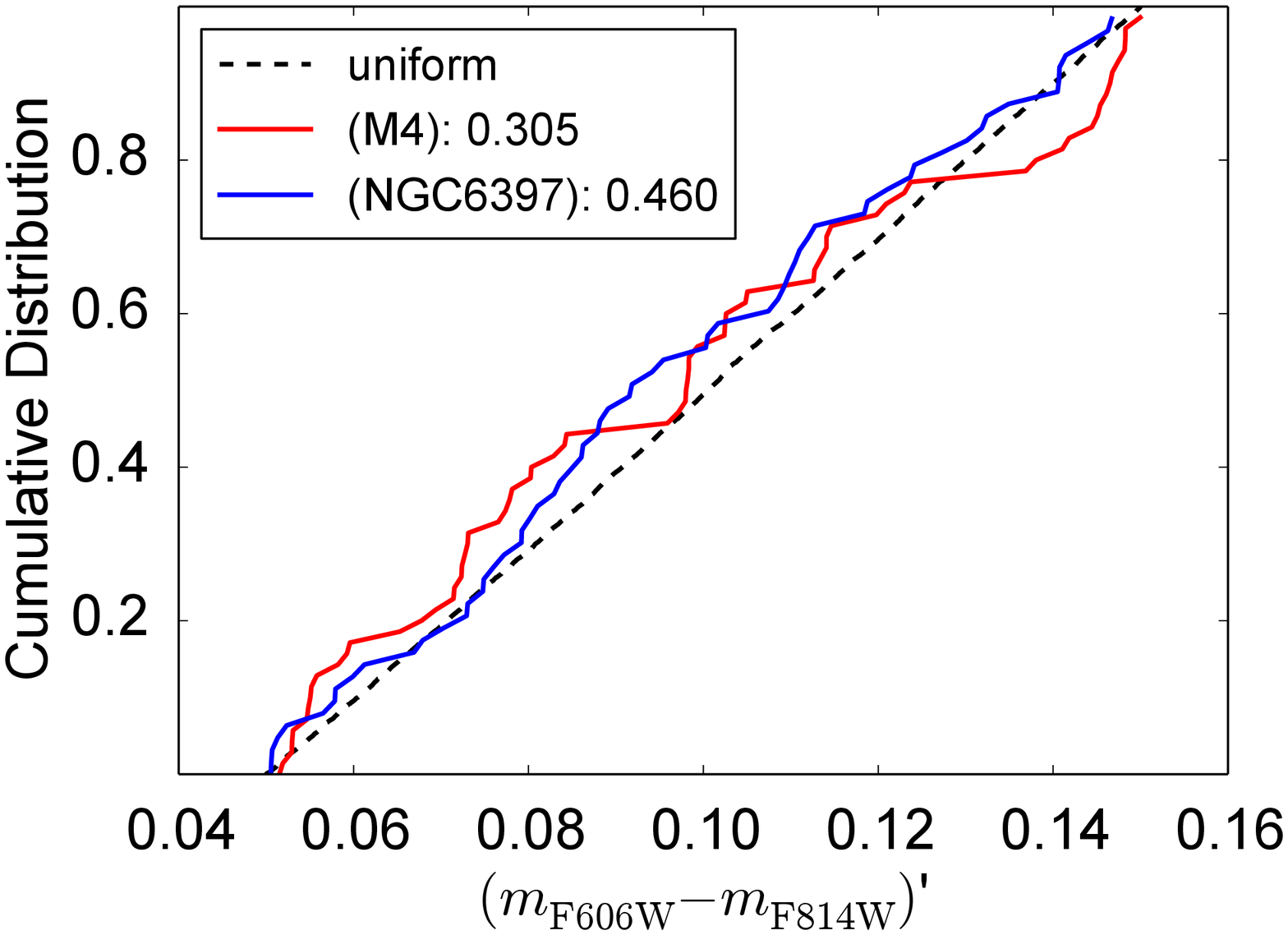} 
    \end{center}
  \caption{CMD (top panels) and colour distributions (middle and bottom panels) 
of M4 (left-hand column) and NGC6397 (right-hand column). 
The CMDs were generated with the HST Globular Cluster Treasury catalogue 
\citep{Sarajedini_2007}, which is based on HST ACS/WFC data, and the
colour distributions were generated following the prescription in Section \ref{problem_model}.
Only stars with $(m_{\rm F606W}-m_{\rm F814W})'>0.05$ are used to compute the above distributions.
The gray areas in the CMDs indicate the region from where the colour distributions 
were derived, and the horizontal lines in the histograms are average numbers, 
assuming flat  distributions.
Notice that in these two GCs, taken as examples because of their proximities, 
pronounced sequences caused by binaries with $q \approx 1$ are not visible, especially in the
region above and close to the turn-off (compare with Fig. \ref{FIG03.1}). 
Additionally, their colour distributions are consistent with uniform distributions. This is
is indicated in the last panel which shows the colour cumulative distributions and the
$p$-values of one-sample Kolmogorov-Smirnov tests for uniformity. The test does not allow us to
reject the hypothesis that they are uniform. For more detail see Sections \ref{problem_real}.
}
  \label{FIG04}
\end{figure}

In order to compare different synthetic CMDs with real CMDs, we evolved
6 models with different initial conditions for 12 Gyr. We adopt for all models the
canonical IMF that follows the broken power law $\xi(m) \propto m^{-\alpha}$, defined by 
\citet{Kroupa_2001}, where $\alpha = 1.3$ for $0.08 \leq m/{\rm M_\odot} \leq 0.5$ and 
$\alpha = 2.3$ for $0.5 \leq m/{\rm M_\odot} \leq m_{\rm max}/{\rm M_\odot}$, 
and the star mass in this study lies between $0.08 {\rm M_\odot}$ and $100 {\rm M_\odot}$.  
Additionally, all models have 95 per cent of primordial binaries\footnote{
We set the initial binary fraction
different from 100 per cent in order to avoid computational problems 
that arise in MOCCA if there is no single star in the initial
model.}. The metallicity equals $Z=0.001$, 
which is typical for GCs, and their binary properties are given by the 
original Kroupa IBP (Section \ref{original}).

We assume that all stars are on the zero-age main sequence
when the simulation begins and that any residual gas from the star
formation process has already been removed from the cluster. Additionally,
all models are initially at virial equilibrium, and have neither
rotation nor mass segregation. Moreover, all models are evolved 
for 12 Gyr which is associated with the present-day in this investigation.

In Table 1, we summarize the main parameters of the initial and present-day
models. Notice that we have a small set of GC models and, even being
so, we still have very different initial and present-day properties, 
which guarantees that a considerable region of the parameter space is covered, 
allowing us in turn to access different CMD morphologies arising from 
different choices of initial conditions.
 
In order to investigate the distribution of binaries in GC models 
simulated with the original Kroupa IBP (Section \ref{original}),
CMDs were obtained by imaging and carrying out photometry on the inner 
parts of the clusters (within their half-light radii) with the {\sc cocoa} code,
assuming that the clusters are at a heliocentric distance of 5 kpc. The observations
were simulated with an HST type telescope with a pixel scale of 0.05''/pix, 
a seeing value\footnote{We emphasize that with HST-like observations we should have no seeing,
a priori. However, in order to perform photometry, we need to achieve a good full width at 
half maximum (FWHM) value, which implies that a definition of seeing is required.} 
in both the filters (B and V) of 0.15'', and a Gaussian PSF was used.
While these idealized mock observations do not model the exact HST PSF, the combination 
of high spatial resolution images with extremely low seeing values can reproduce FWHM 
values for HST images in optical filters. The PSF photometry obtained from 
these synthetic images are comparable to the results from HST photometry particularly 
for the magnitude regime that we are interested in.

From the synthetic CMDs, we obtained the fiducial MS in the following way,
which is similar to the procedure adopted in \citet{Milone_2012}. 
First we select all stars in the magnitude interval of interest, 
which is adopted here as $[5.0,5.5]$. Second we divide such an interval
into 40 bins and obtain the median (in colour and magnitude) in the 2D 
region defined by each interval. Finally the fiducial MS, in the interval, has been 
derived by fitting these median points with a line. From such a fiducial MS, 
we have the line properties, which allows us to rotate and 
translate this region such that the MS becomes a vertical
line centred at the origin of the x axis (the colour axis). From this rotated
and translated part of each CMD, we compute the rotated colour $(B-V)'$ distribution, 
starting from the value $(B-V)' = 0.05$. This lower limit guarantees that the systems,
for $(B-V)' > 0.05$, are likely binaries with $q > 0.5$. System with $(B-V)' < 0.05$
are either single stars belonging to the MS or binaries close to the MS 
(i.e. binaries with $q<0.5$).

Fig. \ref{FIG03.1} exhibits the synthetic CMDs and Fig. \ref{FIG03.2} the colour 
distributions. Notice that the colour distribution is an increasing function of the colour
and that the peaks shown in the right edge of each distribution
are directly connected with the abundant presence of MS binaries whose mass ratios
are $\approx$ 1. The large width of the peak is connected with observational errors, which
are larger for less luminous stars.

From the histograms in Fig. \ref{FIG03.2} we can expect that the distributions are not
uniform, since the dashed horizontal lines (average number, assuming a uniform 
colour distribution) are unlikely good fits for the histograms. In order to confirm
this, we applied a one-sample Kolmogorov-Smirnov test for uniformity to the six 
models. All models exhibit rather small $p$-values (see keys in bottom panel
of Fig. \ref{FIG03.2}). The result of this test allows rejection of the null hypothesis
that the colour distributions are uniform with more than 99 per cent of confidence.
This indicates that, if MS binary mass ratio distributions (and in turn colour distributions) 
in real GCs are consistent with uniformity, then a potential problem with
the Kroupa IBP in matching binary distributions in GCs seems to exist.

In order to show that the peaks in the colour distribution in Fig. \ref{FIG03.2} are instrinsically connected with the 
Kroupa IBP, we show in in Fig. \ref{FIG03.3} the initial and present-day period distributions of
all present-day MS binaries whose $q>0.99$ in all six models. We note that both distributions
are basically the same, which supports that they are
primordial binaries not strongly affected by dynamics. In addition, the majority
of these binaries have periods shorter than 100 days, which makes them very
dynamically hard and not easily subject to disruption via dynamical interactions.

In the following section, we show, as examples, two real GC CMDs and 
their colour distributions, compare them with results obtained for the 
original Kroupa IBP, and conclude that the original Kroupa IBP is probably
not good in matching MS binary distributions in real GCs.


\subsection{Real GC CMDs and the problem with the original Kroupa IBP}
\label{problem_real}

In order to show that GC models set with the 
original Kroupa IBP are not good to reconcile observed
CMD morphologies and MS binary colour distributions, we analyse here
two real GCs, namely M4 and NGC6397. These two clusters are
the nearest GCs \citep[][2010 edition]{Harris} and are also therefore well-studied. 
In addition to the quality of data concerning these two GCs, M4 can be described 
by a classic King profile, while NGC6397 is a post-core collapse cluster, 
which allows us to investigate CMDs and colour distributions in two 
different `dynamical regimes'. 

Fig. \ref{FIG04} exhibits the CMDs and the colour distributions of M4
and NGC6397. The CMDs were generated with the HST Globular Cluster Treasury catalogue 
\citep{Sarajedini_2007}, which is based on HST ACS/WFC data, and the
colour distributions were generated following the prescription in Section 
\ref{problem_model}, being the magnitude interval $[17.5,18.5]$ in both
GCs.

The first incompatibility we notice while comparing Figs. \ref{FIG03.2}
and \ref{FIG04} is the absence of a pronounced binary sequence above the
MS turn-off in the CMDs of M4 and NGC6397. This sort of sequence
is present in all CMDs of Fig. \ref{FIG03.2}, which is caused by 
binaries which are very dynamically hard and have $q \approx 1$, as illustrated
in Figs. \ref{FIG03.2} and \ref{FIG03.3}. Additionally, if 
this feature is present in all CMDs, and if the original Kroupa IBP
should correspond to initial GC binary properties, then such a
behaviour should be also visible in real GCs. Even in the nearest
GCs this is not the case, which implies that the original Kroupa
IBP seems not to be adequate to reproduce binary distributions in real
GCs.

Similarly, contrary to what is predicted when models are set with the
original Kroupa IBP, the colour distributions do seem to be flat in real
GCs, since the dashed horizontal lines in the histograms appear to be a good 
parent model for the distributions. Again, in order to test this in
a more coherent way, we applied one-sample Kolmogorov-Smirnov tests
for uniformity. The $p$-values are 0.305 and 0.46, for M4 and NGC6397,
respectively. This test does not allow us to reject the hypothesis that 
their colour distributions are uniform, which implies that they are consistent
with uniform distributions.

We stress here that we decided to perform the analysis with respect to only two GCs 
because several investigations have already been carried out with regards
to CMDs obtained with HST data \citep[e.g.][]{Piotto_2002,Sarajedini_2007,Milone_2012}.
None of them is consistent with a mass ratio distribution peaked at $q \approx 1$.
In addition, since \citet{Milone_2012} have already concluded that the observed MS mass ratio
distributions in GCs are consistent with uniform distributions, we expect that
our procedure concerning the colour distributions would reveal the same trends found for M4 and 
NGC6397 (i.e. colour distribution consistent with the hypothesis of uniformity), if more
GCs were included.

Additionally, the HST F606W and F814W photometric filter bands approximately correspond to the V and I 
band filters, respectively. So our use of the observational CMDs in terms of 
(V-I) vs I are a very good approximation to HST data sufficient for the present 
purpose. \citet{Askar_2017b} have shown that such a binary sequence is also very 
pronounced (see their figs. 11, 20 and 21) in CMDs generated with the V, U and I 
bands. Therefore, there should not be any serious biases in the compared data sets, 
in particular, since we are not aiming at a detailed modelling of a particular cluster.

Summarising, binary distributions found in present-day GC models set with the original Kroupa IBP 
are unlikely to match those derived from observations and the evidence for that is twofold. First
in part of the CMD above the MS turn-off, a pronounced binary sequence is clearly predicted, although
not observed. Second, the observed colour distributions below the turn-off are consistent with uniform distributions,
and predicted distributions are characterized by a peak in the right edge of the
distributions. As both discrepancies are mainly caused by the presence of a significant fraction of short-period 
low-mass binaries whose component masses are similar, a way to avoid such a non-observed
effect is to reduce the fraction of these systems in the IBP. In what follows, we 
describe how we did this by changing slightly the pre-main-sequence eigenevolution 
prescription as well as what distributions we adopt for high-mass 
binaries, which differ drastically from those of the original Kroupa
IBP.


\section{MODIFIED KROUPA IBP}
\label{modified}

In order to prevent the
discrepancies pointed out in the previous section with respect
to MS binary distributions
(due to binaries whose components
have equal masses) in present-day GC CMDs, we can
change the pre-main-sequence eigenevolution prescription. In what follows,
we describe the procedure to find a revised pre-main-sequence eigenevolution process
as well as what sort of distributions for high-mass binaries we should
adopt in order to respect observational constraints. 


\subsection{Looking for a revised pre-main-sequence eigenevolution}
\label{method}

The first step we adopted in changing the pre-main-sequence eigenevolution is the
strength with which the secondary accretes matter from the circumstellar
disk around the primary. This is intrinsically associated with one 
parameter in Section \ref{latetype1}, namely $\lambda$ (Eq. \ref{EQ4})
and with the efficiency of the accretion expressed by $\rho$ (Eq. \ref{EQ6}).

In order to reduce such a strength of increasing the secondary mass,
we assumed a mass-depend $\lambda$ such that the lower the primary mass,
the weaker the redistribution of energy and angular momentum within their
circumstellar material. This dependence with the primary mass
would take into account differences in binaries with different spectral types.
Thus, we assumed that $\lambda(M_1)$ is a monotonic
function of the primary mass up to a critical primary mass $M_{\rm crit}$ and a constant
after that, i.e.

\begin{equation}
\lambda(M_1) = \left\{
\begin{array}{rcl}
\lambda_0 \times M_1^\eta,& \mbox{if} &  M_1 \leq M_{\rm crit} , \\
\lambda_0           ,& \mbox{if} &  M_{\rm crit} < M_1 < 5 \; {\rm M}_\odot ,\\
\end{array}
\right. \label{EQ8}
\end{equation}
where $\lambda_0$, $\eta$, and $M_{\rm crit}$ are parameters to be determined.

In addition to a mass-dependent $\lambda(M_1)$, we also added a parameter 
$\varepsilon \in (0,1)$ in Eq. \ref{EQ6} (associated with the efficiency at which the
secondary mass accretes matter from the primary circumstellar disk). 
We also added in the expression a stochastic component in the accretion process of 
the secondary for low-mass binaries whose primary masses are smaller than $M_{\rm crit}$.
The modified secondary accretion process, i.e. modified initial mass ratio is given by

\begin{equation}
q_{\rm ini} = \left\{
\begin{array}{rcl}
q_{\rm bir} + \rho_0 (1 - q_{\rm bir})		, & \mbox{if} &  \rho_0 < 1 ,\\
0.9 + 0.1 \times U(0,1)                         , & \mbox{if} &  \rho_0 \geq 1 \ {\rm and} \ M_1 < M_{\rm crit} ,\\
1                                               , & \mbox{if} &  \rho_0 \geq 1 \ {\rm and} \ M_1 \geq M_{\rm crit} ,\\
\end{array}
\right. \label{EQ9}
\end{equation}
where $ U(0,1)$ is the standard uniform distribution and 

\begin{equation}
\rho_0 = \left\{
\begin{array}{rcl}
\varepsilon \rho	, & \mbox{if} &  M_1 < M_{\rm crit} ,\\
\rho			, & \mbox{if} &  M_1 \geq M_{\rm crit} .\\
\end{array}
\right. \label{EQ9.2}
\end{equation}

Note that in the original Kroupa IBP, $\varepsilon = 1$ by construction. Since we aim to reduce
the fraction of short-period low-mass binaries with equal mass components, we allowed $\varepsilon$
(efficiency of secondary mass increase) to be smaller than unity, its value heeding to be determined
as for $\lambda_0$, $\eta$, and $M_{\rm crit}$ . This is a reasonable assumption if one realises
that the change in the eccentricity need not be coupled with the change in the secondary mass. They are
in principle two distinct physical process. In fact, as we will see, the efficiency for changing the
eccentricity is different from that associated with the changes in the secondary mass. 

Note also that we introduced a random initial mass ratio in the range of $(0.9,1.0)$, when
$\rho_0$ is greater than unity and $M_1$ is less than $M_{\rm crit}$.
This was done to account for 
a stochastic component in the accretion process of the secondary.
Again, this improvement was motivated by present-day CMDs which exhibit
pronounced high-q binary sequences
caused by G-dwarfs and K-dwarfs with mass ratio
near 1.0. With such a modification, short-period low-mass binaries unlikely have
components with nearly the same mass.

Finally, note that in this procedure, for convenience, we kept $\chi$ in Eq. \ref{EQ4} fixed 
and given by $3/4$ as in the original Kroupa IBP. We calculated more than 200 models varying 
conveniently the above-mentioned parameters  with the following values

\begin{itemize} 
\item $\lambda_0$: \hspace{0.02cm} 20, 28, 36 ,
\item $M_{\rm crit}$: \hspace{0.02cm} 1.0, 1.5, 2.0 ,
\item $\eta$: \hspace{0.02cm} 0.25, 0.50, 0.75, 1.00, 1.25, 1.50, 1.75, 2.00 ,
\item $\varepsilon$: \hspace{0.02cm} 0.25, 0.30, 0.50.
\end{itemize}

Given each model, we used properties of late-type binaries that are not strongly
affected by dynamics, mainly mass-ratio and eccentricity distribution of short-period binaries,
to compare with observational data of G-dwarf binaries in the Galaxy \citep[][]{DM_1991},
using the minimum $\chi^2$ method of model-fitting. Our best-fit model has the following values:

\begin{equation}
\left\{
\begin{array}{lcl}
\lambda_0    & = &  28 , \\
M_{\rm crit} & = &  1 \ {\rm M}_\odot , \\
\eta         & = &  1/4 , \\
\varepsilon  & = &  1/2 . \\
\end{array}
\right. \label{EQ10}
\end{equation}

Note that $M_{\rm crit} = 1 \ {\rm M}_\odot$ and $\lambda_0 = 28$ 
which implies that our modifications are effective only for G, K and M-dwarfs.
For the rest of the low-mass binaries, the modified Kroupa IBP is exactly
the same as the original Kroupa IBP. 
Note also that $\varepsilon = 1/2$ which implies that the efficiency in the change
of the mass ratio is half in comparison with the change in the eccentricity, for 
binaries whose primary mass is smaller than $M_{\rm crit} = 1 \ {\rm M}_\odot$.
Finally, since $\lambda_0 = 28$ and $\eta = 1/4$,
the $\rho$ as defined in Eqs. \ref{EQ4} and \ref{EQ8} is a smooth, continuous and monotonically
non-decreasing function of the primary mass $M_1$, in the range where pre-main-sequence eigenevolution
should be applied.
 
In Section \ref{results} we show the results for the modified Kroupa
IBP that will be described in the following sections.


\subsection{Low-mass binaries}
\label{latetype2}

To obtain the initial population for low-mass binaries in our revised 
pre-main-sequence eigenevolution procedure, we follow exactly the same approach as that 
to generate the original Kroupa IBP. The only difference in the modified
Kroupa IBP is the different equations that are used.

We first generate the birth population with distributions (i--iv) in
Section \ref{birth1}. Then we convert the birth population to the
initial population using Eqs. \ref{EQ8}, \ref{EQ4}, \ref{EQ5}, 
\ref{EQ9}, and \ref{EQ7}, with the parameter values given by Eq. \ref{EQ10}.

In Fig. \ref{FIG01} we plot the main distributions associated with the modified Kroupa 
IBP for low-mass binaries, i.e. primary mass (top left-hand panel), mass ratio 
(top right-hand panel), period  (bottom left-hand panel), and eccentricity (bottom 
right-hand panel). Note that the original Kroupa IBP (Section \ref{latetype1}) 
was included for comparison.

Note that the main difference between the original and the modified Kroupa IBPs
is the mass ratio distribution. 
Indeed, in the modified Kroupa IBP the fraction
of binaries in the range of $q \in [0.9,1.0]$ is reduced due to the revised $\lambda$ 
(Eq. \ref{EQ8}) and the revised secondary accretion process (Eq. \ref{EQ9}).


\subsection{High-mass binaries}
\label{earlytype2}

In contrast with the original Kroupa IBP, for high-mass binaries,
we adopt observational distributions derived for O and B-dwarfs
to generate the birth population. Similarly to the original Kroupa
IBP, no pre-main-sequence eigenevolution is applied in our modified version.

We adopt the distributions derived by \citet{Sana_2012} who
analysed the O star population of six nearby Galactic open
stellar clusters. Note that the binarity found by \citet{Sana_2012}
is not 100 per cent and, since we assume, as usual, that star clusters
are formed with a dominant population of binaries, i.e. with $\approx$ 100 
per cent of binaries, we should extend those distributions and normalise them
in order to take this into account.

Following \citet{Oh_2015}, the period distribution is given by

\begin{equation}
f_P \ = \ 0.23 \times \log_{10}(P/{\rm days})^{-0.55} , \label{EQ11}
\end{equation}
where $\log_{10}(P/{\rm days}) \in [0.15,6.7]$. This distribution and
range ensure that the cumulative binary fraction becomes unity.

In an analogous way, we extend the eccentricity distribution up
to unity which results in the following distribution

\begin{equation}
f_e \ = \ 0.55 \times e^{-0.45} , \label{EQ12}
\end{equation}
where $e \in [0.0,1.0]$.

Regarding the mass ratio, \citet{Sana_2012} found a uniform distribution, i.e.

\begin{equation}
f_q \ = \ 1.0 , \label{EQ13}
\end{equation}
where $q \in [0.1,1.0]$.

We emphasize that the correct way of pairing birth low-mass and high-mass binaries and preserving
the IMF is as follows. First, we generate an array of all stars (twice the number of binaries)
from the IMF.  Second, we select the more massive star (primary) from the array of stars, a mass 
ratio from the uniform distribution, and then compute the `ideal' secondary mass. After that, 
the star that has the closest mass to the `ideal' secondary mass is chosen from the array to be paired with
the primary. Finally, both stars are removed from the array. Then, we proceed with the binary
generation until the primary mass is smaller than 5 M$_\odot$. After that point, the above procedure
remains the same with the exception that the secondary is chosen randomly in the array. 

Finally, as done by \citet{Kroupa_1995b} for low-mass binaries,
the upper envelope and circularization period in the plane $e$ vs. 
$\log_{10}(P)$ for high-mass binaries have to be consistent with
to observational data. Here, we applied the criterion derived in 
\citet{Moe_2016} who considered observational data of O and B-dwarfs.
This guarantees that the binary components  do not fill their Roche-lobe 
by a factor greater than 70 per cent at the pericentre and it is achieved 
by the following restriction

\begin{equation}
e_{\rm max} \ = \ 1 \ - \ \left( \frac{P}{P_{\rm circ}} \right)^{-2/3} , \label{EQ14}
\end{equation}
where $P > P_{\rm circ}$, and the circularization period is given by
$P_{\rm circ} = 2$ days.

In Fig. \ref{FIG02}, we show the main distributions associated with the modified Kroupa
IBP with respect to high-mass binaries, i.e. primary mass (top left-hand panel), mass 
ratio (top right-hand panel), period  (bottom left-hand panel), and eccentricity 
(bottom right-hand panel). We also include the distributions of the original Kroupa IBP 
(Section \ref{earlytype1}) for comparison, as in Fig. \ref{FIG01}.

We emphasize here that a way to determine the birth high-mass binary
distributions would be to investigate if the assumed hypothesis leads 
to the observed number of runaway massive stars through dynamical mass 
segregation to the cluster core and partner exchanges through
dynamical encounters there between the massive stars \citep{Kroupa_INITIAL}.
This was partially done by \citet{Oh_2016} who showed that the birth mass 
ratio distribution for O-star primaries must be near uniform for mass 
ratios greater than $0.1$, consistent with observational results of
\citet{Sana_2012}. Finally, as massive stars require their own process of
converting the birth population into the initial population, 
we assume here that such a process has already taken place and
we adopt directly observational distributions, which is a reasonable approach
considering the lack of understanding in this process. This is
also consistent with the fact that this process should be very fast for massive
stars \citep{Railton_2014}.


\section{RESULTS}
\label{results}

In this section, we present the main results associated with the
modified Kroupa IBP in a comparative way with the original Kroupa
IBP. We show results for Galactic field late-type binaries and
GCs. We end this section by showing that our modified 
Kroupa IBP solves the problem associated with present-day GC CMDs.


\subsection{Stimulated evolution}
\label{stimulated}

In order to compare our results with Galactic
field binaries, we assume as usual that Galactic field binaries
come from the dissolution of dense star clusters, after stimulated evolution \citep{Kroupa_1995b}.
Such a dissolution is assumed to be caused mainly by the residual
gas removal due the evolution of massive stars (winds and, eventually, supernovas). 

For the embedded clusters, we assume that initial clusters follow 
the Marks-Kroupa relation between half-mass radius and embedded mass of stars 
in the embedded cluster \citep{Marks_2012}, i.e.

\begin{equation}
\left( \frac{R_h}{{\rm pc}} \right) \ = \ 0.1 \, \left( \frac{M_{\rm ecl}}{{\rm M}_\odot} \right)^{0.13} , \label{EQ15}
\end{equation}
where $R_h$ is the cluster initial half-mass radius and $M_{\rm ecl}$
is the stellar mass in the embedded cluster.

Even though this is the most accurate approach, this is not doable
yet with the current version of the {\sc mocca} code, because the expansion
of the embedded cluster due to residual gas removal, which takes place after few $10^5$ yr 
of cluster evolution \citep[e.g.][]{Banerjee_2017}, is not yet implemented in {\sc mocca}.

In order to mimic this expansion and residual gas removal, 
we performed many numerical experiments to look for an initial expansion 
factor in the Marks-Kroupa relation to provide good agreement with observations. 
In order to find the best expansion factor, we assumed five different values, 
namely 1, 2, 5, 7, and 10, that should be multiplied by the half-mass radius in
Eq. \ref{EQ15} to provide appropriate input for {\sc mocca} models. In addition,
since the stimulated evolution varies with the cluster mass, we also verify which
expansion factor gives better results for combinations of stimulated evolution
timescales and cluster mass. We assumed three different timescales, namely 1, 5 and 10 Myr,
and four different cluster masses, which is given in {\sc mocca} by the number of objects
(single stars + binaries), 
namely 10k, 20k, 50k and 100k. Note that these different timescales and
cluster masses allow us to infer the strength of the dynamical processing in 
shaping the distributions through stimulated evolution.

All these models were evolved and compared with the observational mass-dependent binary
fraction \citep[e.g.][their Fig. 2]{Dorval_2017} and properties of Galactic field late-type 
binaries (Section \ref{gflatetype}).

The best agreement with observations was
achieved for an expansion factor of $\approx$ 2 -- 4, for a stimulated evolution 
timescale of $\approx$ 5 Myr. This is also consistent with the results found by \citet{Pelupessy_2012} 
and \citet{Brinkmann_2016} who showed, using direct $N$-body calculations, that embedded clusters expand by approximately 
this factor, if the star-formation efficiency is $\approx$ 1/3, i.e. if $\approx$ 2/3 of the initial gas 
is lost in the star cluster formation process.
The properties of this MOCCA model are given in Table \ref{TabMOCCA}.

\begin{table} 
\caption{Initial conditions of the best-fit model, which was used in the stimulated evolution,
obtained by averaging over all 10 models. See Section \ref{stimulated}, for more details.}
\label{TabMOCCA}
\begin{adjustbox}{max width=530px}
\noindent
\begin{threeparttable}
\noindent
\begin{tabular}{l|l}
\hline\hline
Mass [${\rm M}_\odot$]		 			& $(9.23 \pm 0.16) \times 10^3$	\\
Number of objects		 			& $10^4$	 		\\
Binary fraction\tnote{a} \, [\%]		 			& $95$		 		\\
Metallicity	[Z$_\odot$]		 		& $0.02$	 		\\
Central density	[${\rm M}_\odot \; {\rm pc}^{-3}$]	& $(1.5  \pm 0.5) \times 10^4$ 	\\
Tidal radius [pc]					& $26.0$ 			\\
Half-mass radius [pc]		 			& $0.86 \pm 0.01$ 		\\
Core radius [pc]		 			& $0.28 \pm 0.01$ 		\\
Galactocentric distance [kpc]	 			& $8.01 \pm 0.07$ 		\\
Half-mass radius relaxation time [Myr]			& $32.5 \pm 0.7$  		\\
Central velocity dispersion [km s$^{-1}$]		& $5.1  \pm 0.7$ 		\\
Duration of stimulated evolution [Myr]			& $4.9  \pm 0.5$  		\\
\hline\hline
\end {tabular}

\begin{tablenotes}
       \item[a] We set the initial binary fraction different from 100 per cent in order \\
to avoid computational problems that arise in MOCCA if there is no \\
single star in the initial model.
\end{tablenotes}

\end{threeparttable}
\end{adjustbox}
\end{table}

Assuming an expansion factor of $\approx$ 3, the initial half-mass radius of {\sc MOCCA} models should be

\begin{equation}
\left( \frac{R_{h,\rm MOCCA}}{{\rm pc}} \right) \ \approx \ 0.3 \, \left( \frac{M_{\rm ecl}}{{\rm M}_\odot} \right)^{0.13} , \label{EQ16}
\end{equation}

Note that \citet{Marks_2011} found that 50 per cent of all Galactic field binaries
originate from clusters with $M_{\rm ecl} \lesssim$ 300 M$_\odot$. This in principle
might be a problem due to the fact that {\sc mocca} models should contain at least
10k objects (single stars + binaries) which gives, in general, masses of one order of magnitude higher than
that. However, in our experiments, models with 10k and 20k give similar results
and the 10k model might be used as an upper limit associated with the influence of dynamics
during stimulated evolution. Said that, we decided to show the results for
a cluster of 10k which gives not only such an upper limit, but also is the
closest possible {\sc mocca} model to a low-mass embedded cluster from which the majority
of Galactic field binaries originate. 
The validity of this assertion is demonstrated in the Appendix, in which we show that
during the phase of the cluster expansion connected with residual gas
removal, binaries with periods longer than $\approx 10^{7}$ days are modified by
dynamical interactions. Therefore, only the tail of the observed period distribution can be
modified by dynamics before the binaries become a part of the field population.

{\it This {\sc mocca} model is then dynamically equivalent for the binary population to the embedded clusters which have an expansion phase.}
We define `{\it dynamical equivalence}' as follows: if two clusters with different masses and 
different initial radii dynamically evolve an identical initial binary population to 
similar distribution functions of binaries, then these two clusters are `dynamically 
equivalent'. Cluster evolution need not be over the same time scales for dynamical 
equivalence to be reached. 

For instance, when comparing the results for the models in 
\citet{Kroupa_1995b}, composed initially of 200 binaries and computed for a few hundred Myr, with the models in 
\citet{Kroupa_2001} composed of 5,000 binaries and integrated for a Myr, we notice that both clusters 
are dynamically equivalent. The corresponding binary distribution functions for these two clusters look 
very similar. They started with the same binary-star distributions and were evolved in very different 
cluster environments with and without stellar evolution (compare, for instance, fig. 7 in \citet{Kroupa_1995b}
 with figs. 10 and 11 in \citet{Kroupa_2001}). In addition, \citet{Giersz_2016} showed 
that {\sc mocca} and N-body model binary distributions are evolved in a remarkably similar way, not only their 
period distributions (see their Fig. 2 which shows a comparison between {\sc mocca} model and {\sc BiPoS} code 
\citep{Marks_2011}), but also for binding energy, mass ratio and eccentricity distributions. 

We emphasize 
here that dynamically equivalent models do not have exactly the same binary populations (i.e. same binary 
distributions) after stimulated evolution. However, they are very similar and with this concept it is 
possible to find general solutions to the class of dynamically equivalent clusters which lead to a very 
similar final binary population after finding one N-body solution (sec. 4.3 and sec. 6.4 in \citet{Kroupa_1995a} 
and sec. 4 in \citet{Kroupa_1995c})\footnote{ 
We would like to emphasize that using dynamical equivalence does not imply that a cluster evolves dynamically in the same way as another cluster. It merely means that, over a certain time scale, the dynamical evolution processes the binary population similarly.}.

In order to further illustrate this important concept, we show in the
Appendix \ref{ap2}, as an example, that three different initial {\sc mocca} models are dynamically
equivalent over different stimulated evolution timescales by showing their resulting binary distributions
 (Fig. \ref{FIG_AP})\footnote{
We note that a demonstration of dynamical equivalence as defined here in the form of low-N dynamical simulations for the specific case studied in this work will be useful to further test this conjecture, whereby the small-N computations presented in \citet[][]{Kroupa_1995a,Kroupa_1995b,Kroupa_1995c} are the basis of formulating it. There, the scaling from one dynamically-equivalent solution to another is also discussed.}. 

All the above implies that our approach is reasonable and that 
{\sc mocca} models represent clusters which have a dynamically equivalent history to the real embedded clusters 
which have undergone an expansion phase.

The stimulated evolution was performed as follows. 
We evolved for $\approx$ 5 Myr two star cluster models with $M_{\rm ecl}$ 
$\approx 10^4$ M$_\odot$ and 95 per cent of binaries that follow the original
Kroupa IBP and the  modified Kroupa IBP. For each cluster model, we performed
10 realizations in order to reduce fluctuation that arise from the small number
of binaries in the clusters ($\approx$ 10k).

In what follows, we show the results for our averaged (over 10 realizations)
above-described models following Eq. \ref{EQ16}.


\subsection{Galactic field late-type star distributions}
\label{gflatetype}

\begin{figure}
   \begin{center}
    \includegraphics[width=0.49\linewidth]{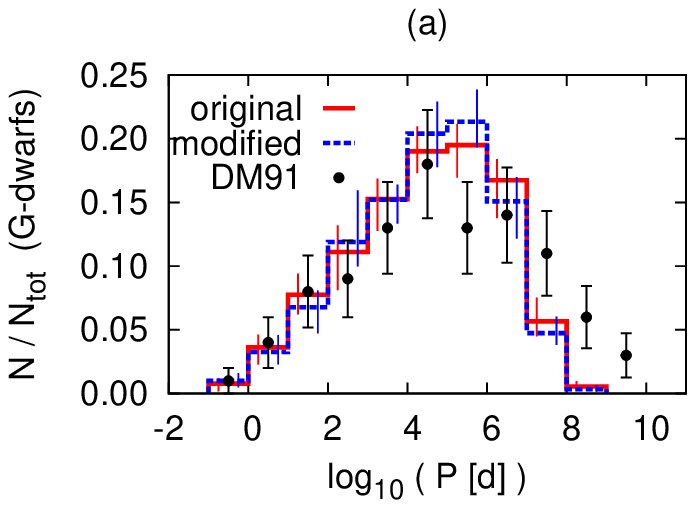} 
    \includegraphics[width=0.49\linewidth]{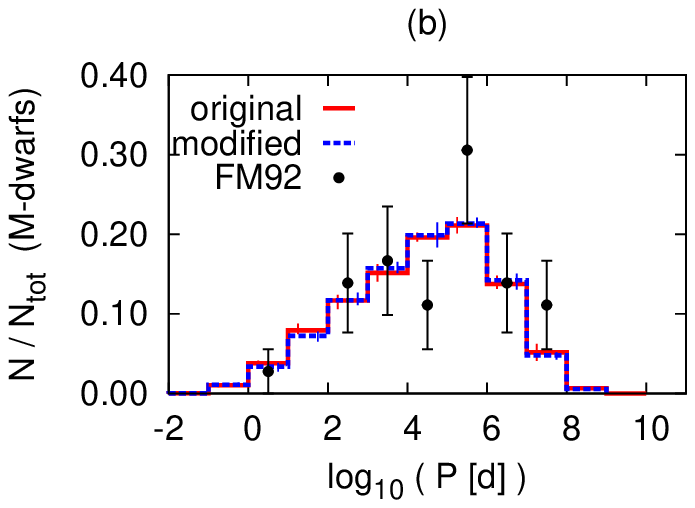} 
    \includegraphics[width=0.49\linewidth]{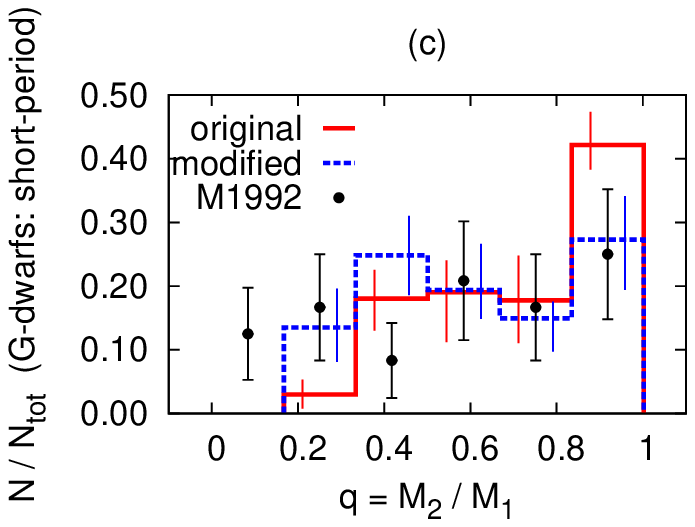} 
    \includegraphics[width=0.49\linewidth]{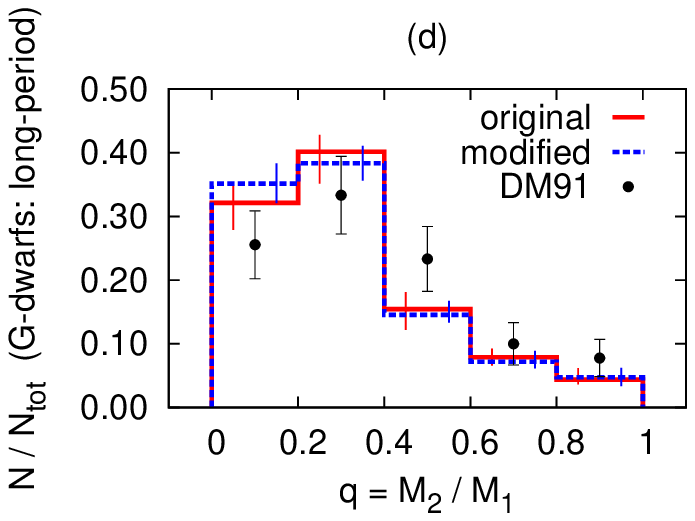} 
    \includegraphics[width=0.49\linewidth]{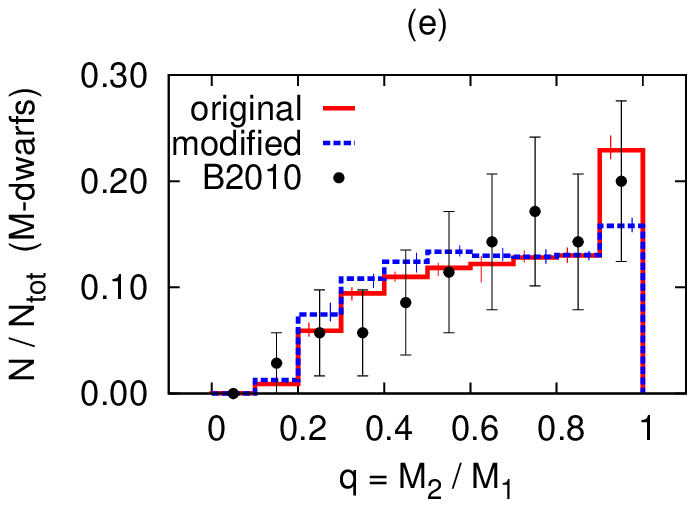} 
    \includegraphics[width=0.49\linewidth]{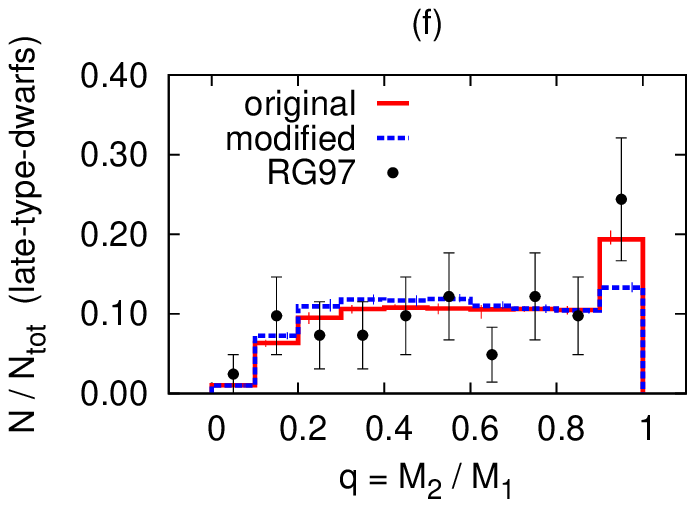} 
    \includegraphics[width=0.49\linewidth]{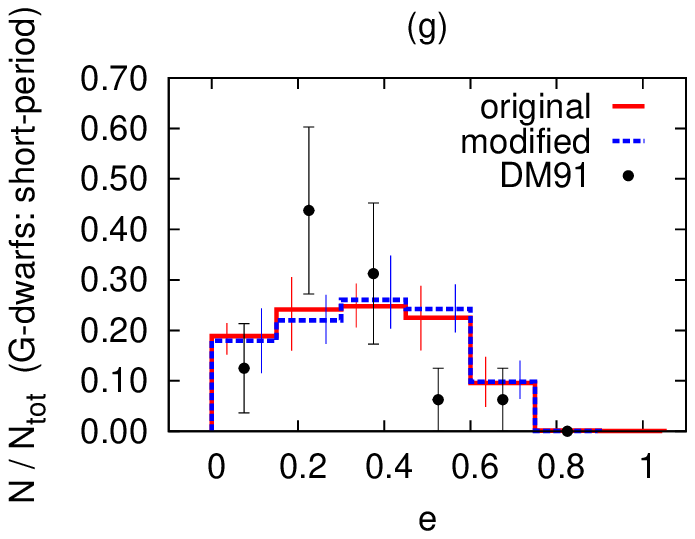} 
    \includegraphics[width=0.49\linewidth]{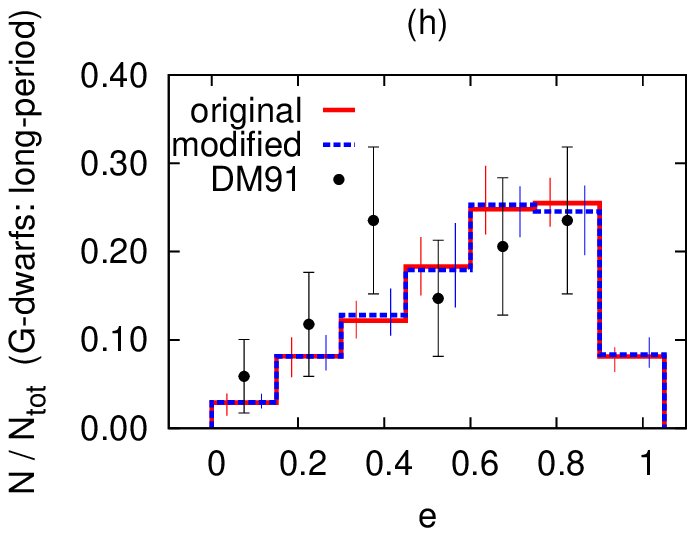} 
    \end{center}
  \caption{Comparison with Galactic field late-type binaries, after stimulated evolution, 
for an averaged model evolved with the original Kroupa IBP (Section \ref{original}, solid line) and 
an averaged model evolved with the modified Kroupa IBP (Section \ref{modified}, dashed line). 
In panels (a) and (b), we plot the period distributions of G and M-dwarfs respectively. 
In panels (c) and (d) we show the mass ratio distributions of short-period and long-period
G-dwarfs, respectively. Mass ratio distributions for M-dwarfs and all dwarfs such that 
$M_1$ is smaller than 2 M$_\odot$ are shown in panels (e) and (f), respectively. Finally,
in panels (g) and (h), we exhibit the eccentricity distributions for short-period and
long-period G-dwarfs, respectively. Observational data is plotted with filled circles and
were extracted from \citet{DM_1991} (DM91), \citet{M_1992} (M1992), \citet{FM_1992} (FM92), 
\citet{RG_1997} (RG97), and \citet{B_2010} (B2010). Vertical lines for the IBPs correspond to 
the entire range of all 10 realizations outcomes, and for the data to Poisson errors. 
For more details, see Section \ref{gflatetype}.}
  \label{FIG05}
\end{figure}

One of our goals, apart from the solution of the problems presented in Section \ref{problem} related
to GC CMDs, is to not worsen the pre-main-sequence eigenevolution process with respect to late-type binaries in the Solar 
neighbourhood, i.e. eccentricity, period and mass ratio correlations, and for short-period binaries, a 
bell-shaped eccentricity distribution and a mass ratio distribution which rises towards unity.

These observational features were derived from long-term observations and here we compare both original
and modified Kroupa IBPs with distributions extracted from the following studies\footnote{Our 
definitions for the late-type binaries are as follows: all late-type, G and M-dwarfs have primary
masses in the ranges $[0.08,2.0]$, $[0.8,1.2]$, and $[0.08,0.6]$, respectively, all in units of M$_\odot$.}:

\begin{itemize} 
\item all G-dwarf period: DM91 \citep[][]{DM_1991},
\item all M-dwarf period: FM92 \citep[][]{FM_1992},
\item all G-dwarf eccentricity: DM91,
\item short-period G-dwarf mass ratio: M1992 \citep[][]{M_1992},
\item long-period G-dwarf mass ratio: DM91,
\item all M-dwarf mass ratio: B2010 \citep[][]{B_2010},
\item all late-type-dwarf mass ratio: RG97 \citep[][]{RG_1997}.
\end{itemize}

Note that we compare the mass ratio distribution of M-dwarf binaries with the data from
\citet[][]{B_2010}, that provide a similar distribution to that from \citet[][]{FM_1992}.

For G-dwarfs, we opted for not including the data from 
\citet[][]{R_2010} for the following reason. As already pointed out by \citet{Marks_2011b},
\citet[][]{R_2010} found a mass-ratio distribution for binaries with a solar-type primary 
that is different from \citet[][]{DM_1991}. Since the range of primary masses
in both works are similar, their mass-ratio distributions should be also similar if 
the periods are comparable. The reason for that is likely due to different approaches
in both works. The survey performed by \citet[][]{DM_1991} lasted 13 years and they
found accurate orbital solutions for their binaries. On the other hand, \citet[][]{R_2010} 
compiled data from different sources and techniques. Given that, we preferred
to compare our models with data from \citet[][]{DM_1991}\footnote{See also 
the comments in \citet{Kroupa_2009}.}.

Additional observational works could be also included in comparison.
For instance, the results by \citet{Halbwachs_2003} who investigated early-type F- and K-dwarf binaries
and those by \citet{Fisher_2005} who presented an incomplete survey of spectroscopic binaries. 
In order to use those results, one could apply the Kroupa IBP to this data set using stimulated evolution, 
but only by also modelling the complicated biases and selection effects, which is, however, out of scope of
the present investigation. Finally, \citet{Rastegaev_2010} presented a survey of population II (sub-dwarf) 
stars. The authors note that the population II period distribution is narrower and 
biased towards shorter periods than the population I 
distribution (their Fig. 10). Even though analysing population II stars is not an objective here,
it will be interesting to assume the universality hypothesis and then to investigate which population 
of star clusters may, if at all, account for the observed population II binary distributions. 
The mass function of metal-poor (population II) star clusters may have been different to that observed 
today for population I. In such an investigation, it will be necessary to model all selection effects and 
biases inherent to the analysis by \citet{Rastegaev_2010}. A particular aspect of such a modelling will be 
the inclusion of population I stars and binaries which have acquired large proper motions,
e.g. by being ejected out of their population I birth clusters. Such a contribution is likely to be found 
in a proper-motion selected sample in addition to population II, and may skew or alter the deduced binary 
star distribution functions. Again, modeling population II via the Kroupa IBP and stimulated evolution 
is not in the scope of this investigation, but is doable in future works.

In Fig. \ref{FIG05}, we compare the observational distributions used in this work with our averaged models,
after stimulated evolution, concerning the original Kroupa IBP and the modified Kroupa IBP.
Vertical lines in the figure correspond to the entire range of all 10 realizations outcomes 
(for the IBPs), and to Poisson errors (for observational data).

Note that both IBPs provide good agreement with observations. In addition, both IBPs give
similar distributions with the exception of the mass ratio distributions.

In panel (c), we show the mass ratio distribution of short-period binaries whose primaries
are G-dwarfs. Note that the fraction of binaries with high mass ratio $q \gtrsim 0.9$ is reduced,
this is expected given our initial motivation for changing the pre-main-sequence eigenevolution prescription.

In panel (e) and (f) we show mass ratio distributions of all binaries whose primaries are
M-dwarfs and late-type-dwarfs, respectively. As for G-dwarfs, we also see a reduction in the 
fraction of high mass ratio binaries. This is again expected due to the revised pre-main-sequence eigenevolution
derived here.

Finally, Fig. \ref{FIG05} clearly shows that none of the previous results achieved with the
original Kroupa IBP are damaged when our modified Kroupa IBP is adopted. In other words,
our modified Kroupa IBP is not only consistent with observational data of 
late-type binaries in the Galactic field, but also provides
qualitatively similarly good description when compared with the original Kroupa IBP.


\subsection{Globular cluster colour-magnitude diagrams}
\label{gccmd}

\begin{figure}
   \begin{center}
    \includegraphics[width=0.48\linewidth]{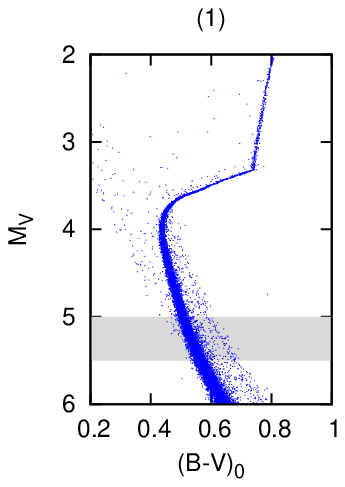}
    \includegraphics[width=0.48\linewidth]{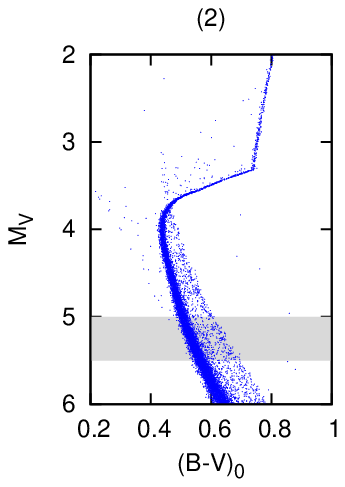}
    \includegraphics[width=0.48\linewidth]{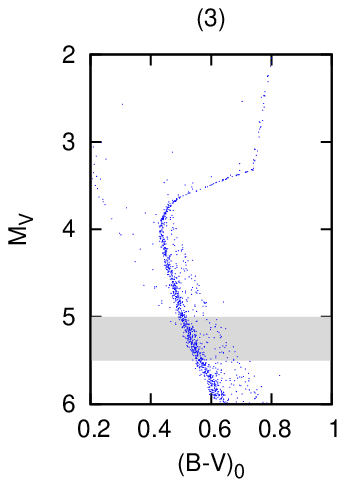}
    \includegraphics[width=0.48\linewidth]{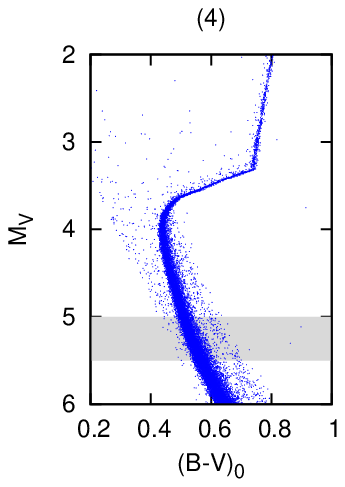}
    \includegraphics[width=0.48\linewidth]{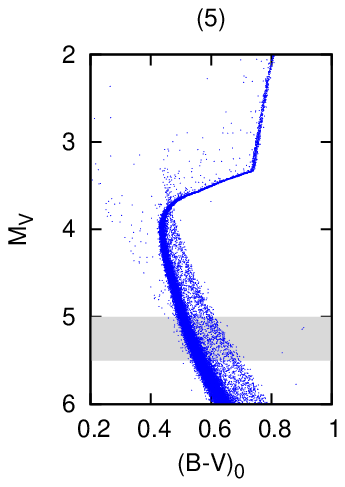}
    \includegraphics[width=0.48\linewidth]{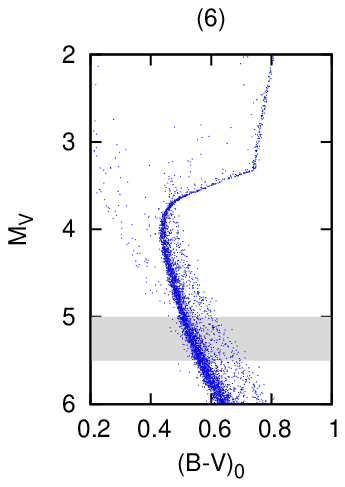}
    \end{center}
  \caption{Present-day CMDs from simulated observations of the six GC models 
(Table 2) evolved with the modified Kroupa IBP (Section \ref{modified}). 
The gray areas indicate the regions used to generate the colour distributions in Fig. \ref{FIG06.2}.
Note that there is no clear and pronounced binary sequence due 
to short-period low-mass binaries with mass ratios of unity in such CMDs,
specially close to the turn-off. 
For more detail see Section \ref{gccmd}.
}
  \label{FIG06.1}
\end{figure}

\begin{figure}
   \begin{center}
    \includegraphics[width=0.49\linewidth]{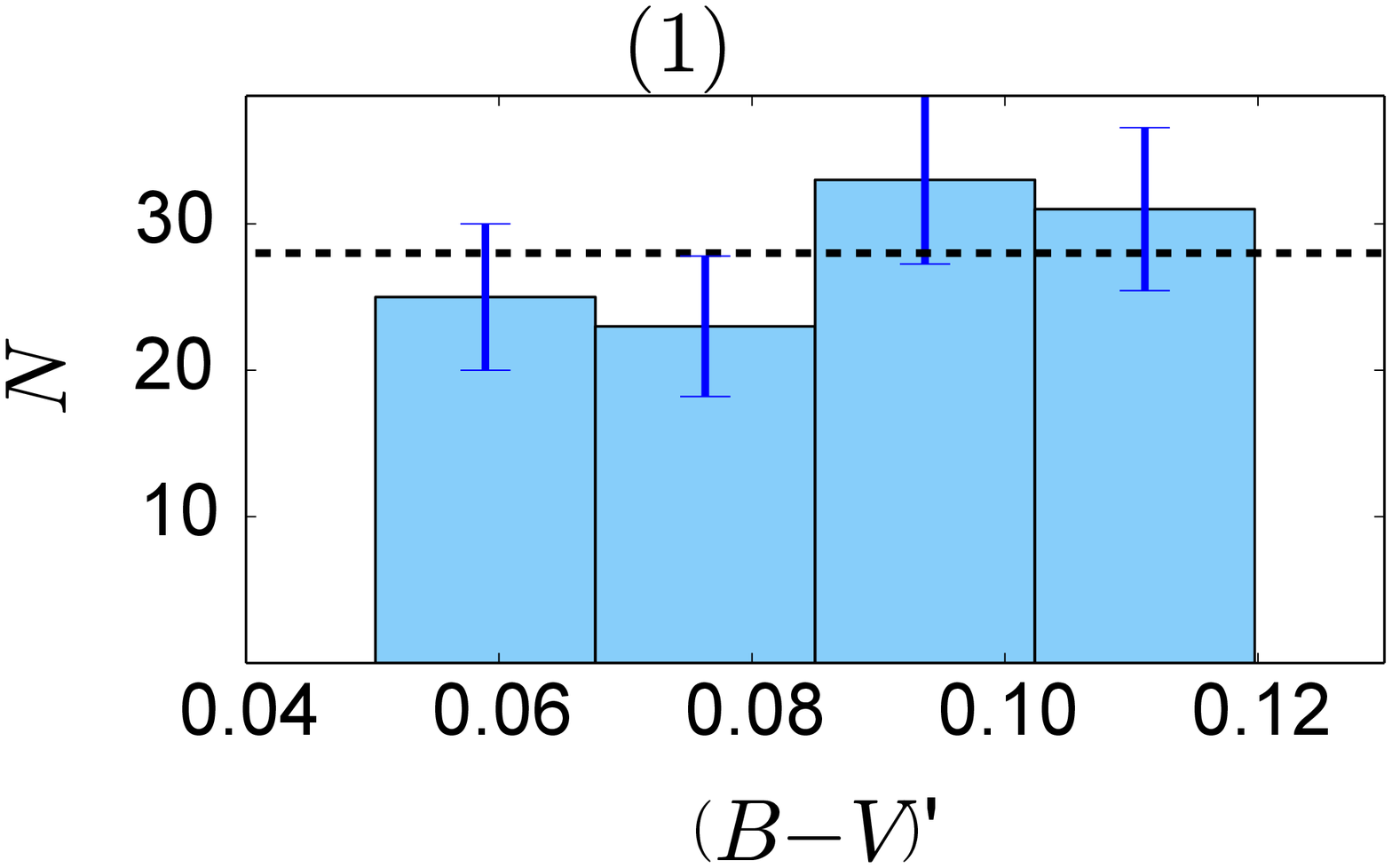} 
    \includegraphics[width=0.49\linewidth]{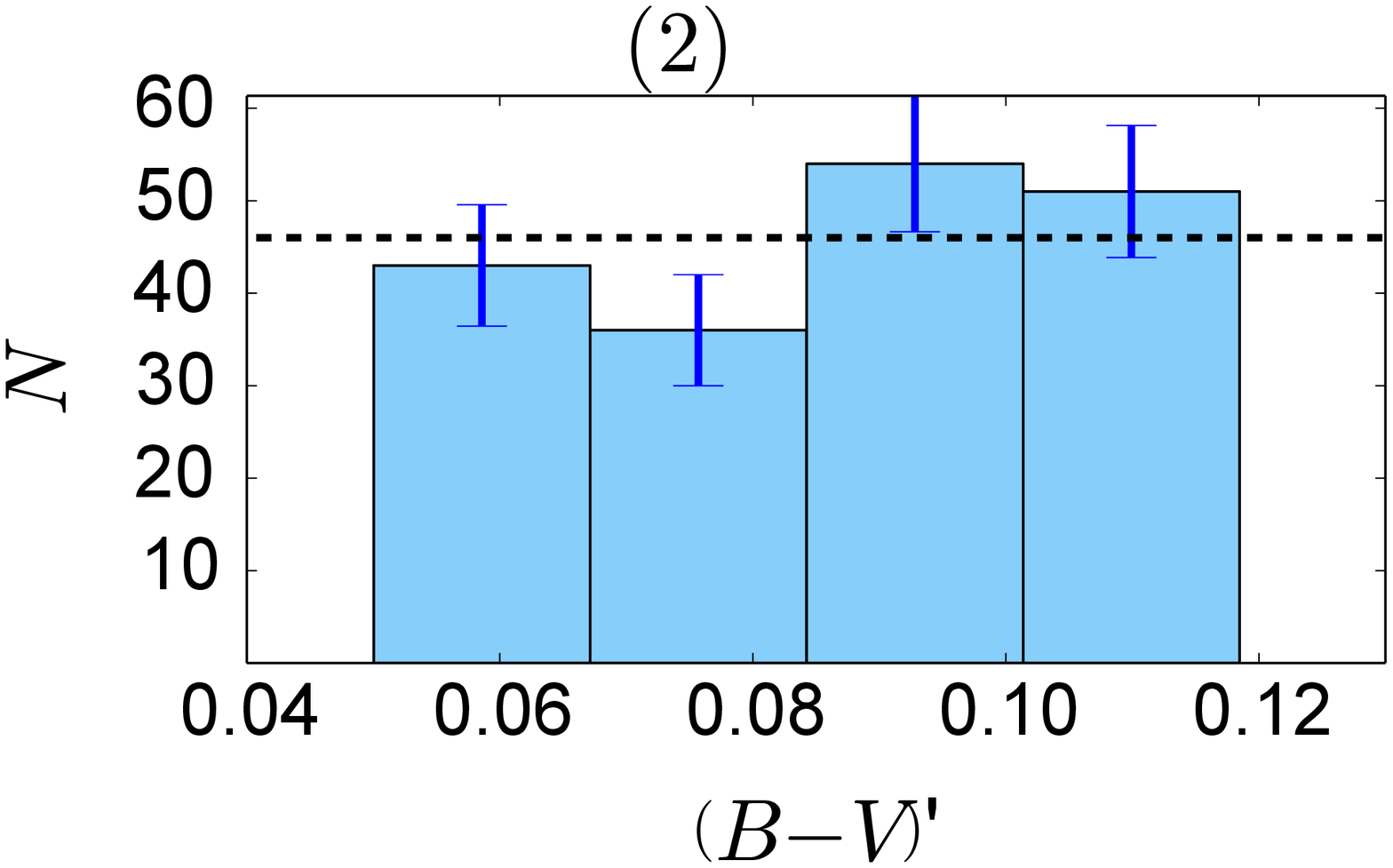} 
    \includegraphics[width=0.49\linewidth]{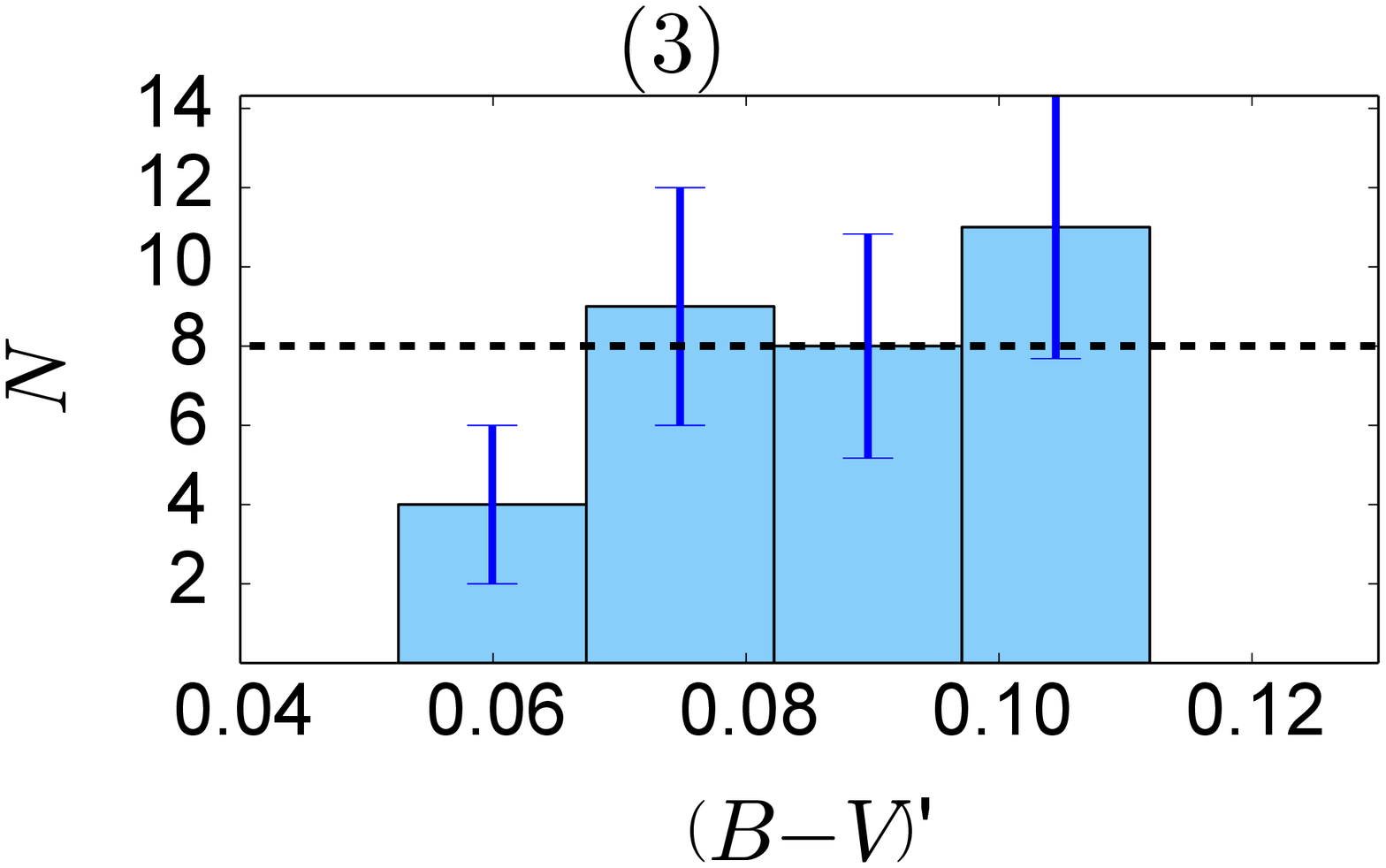} 
    \includegraphics[width=0.49\linewidth]{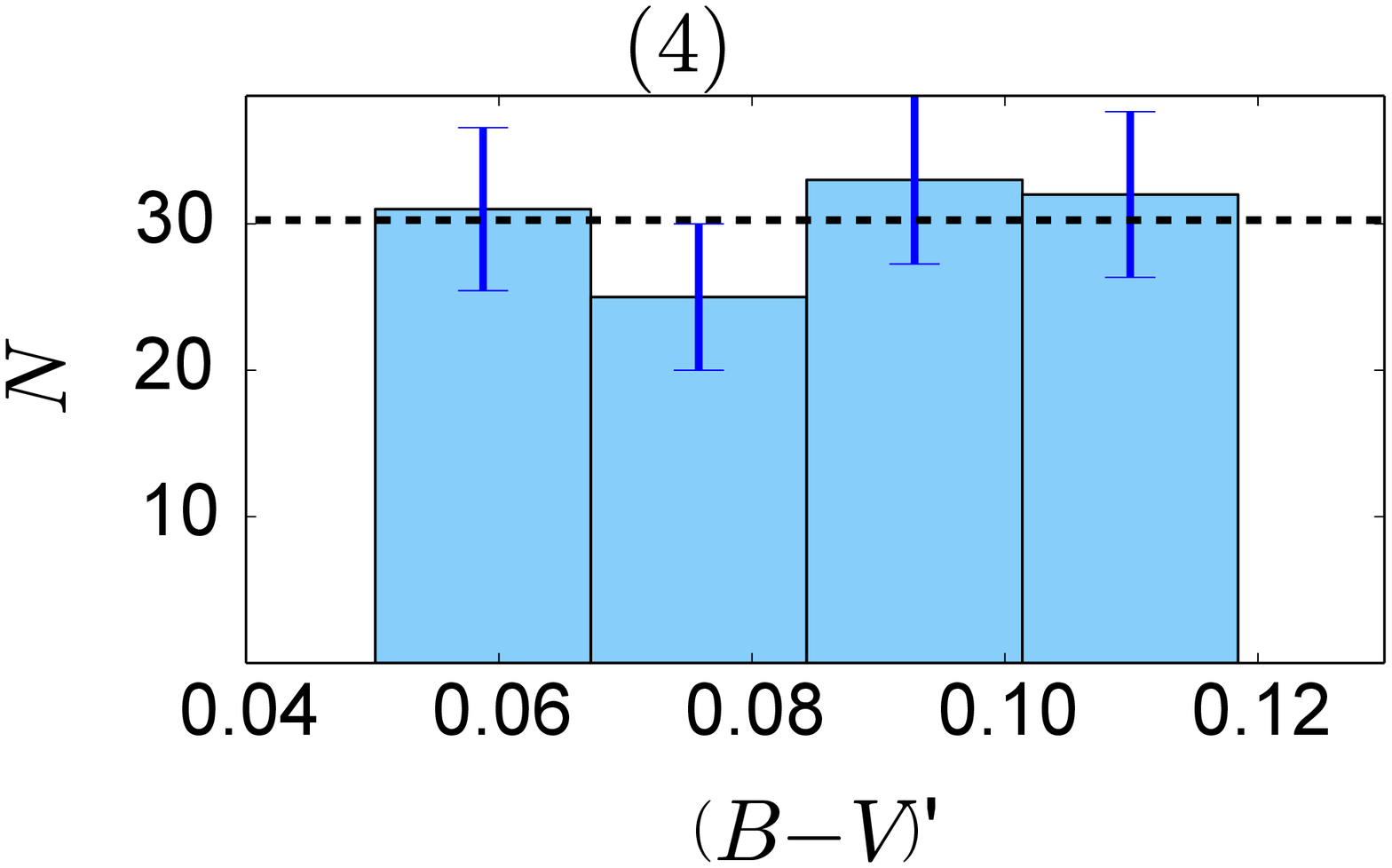} 
    \includegraphics[width=0.49\linewidth]{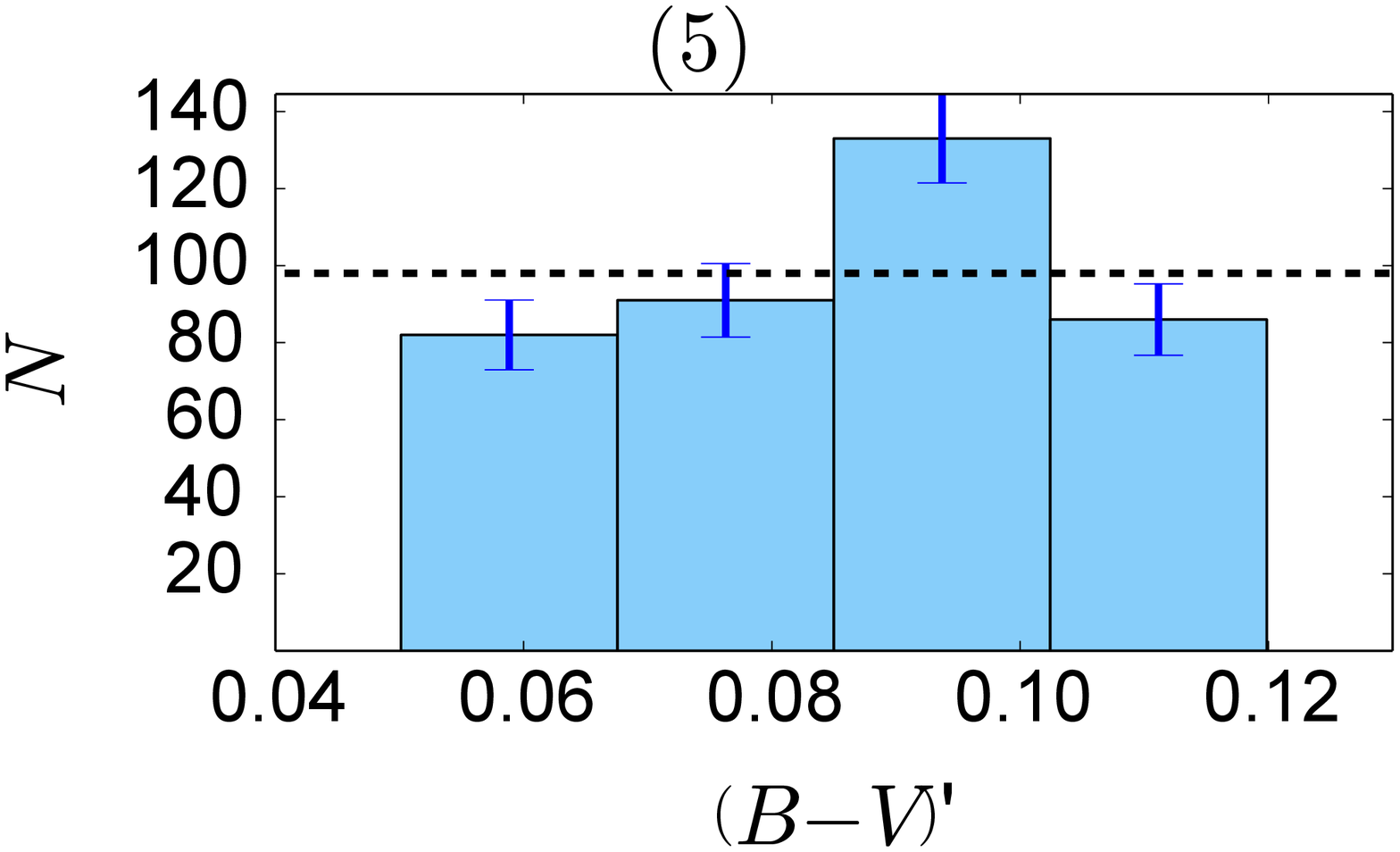} 
    \includegraphics[width=0.49\linewidth]{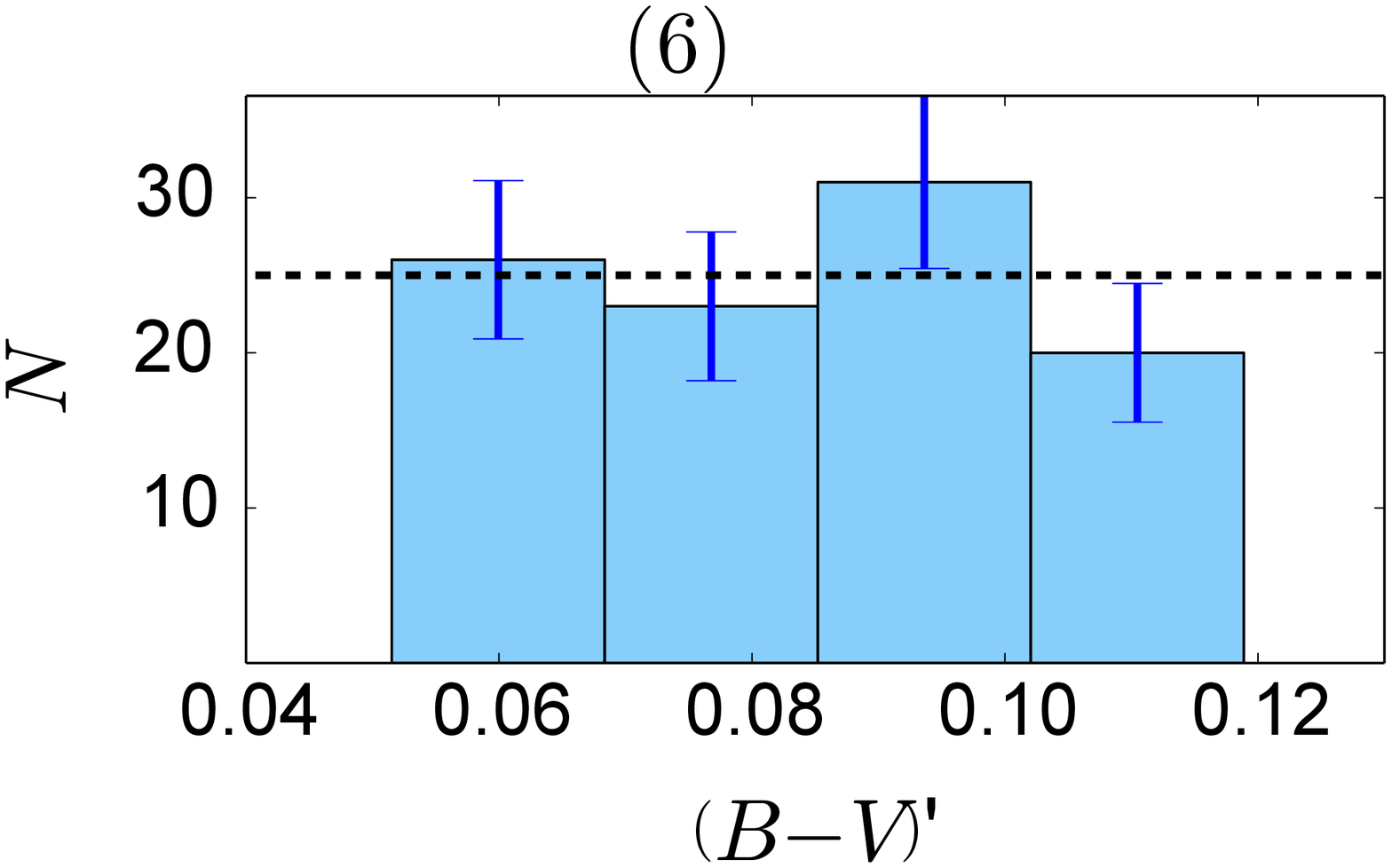} 
    \includegraphics[width=0.95\linewidth]{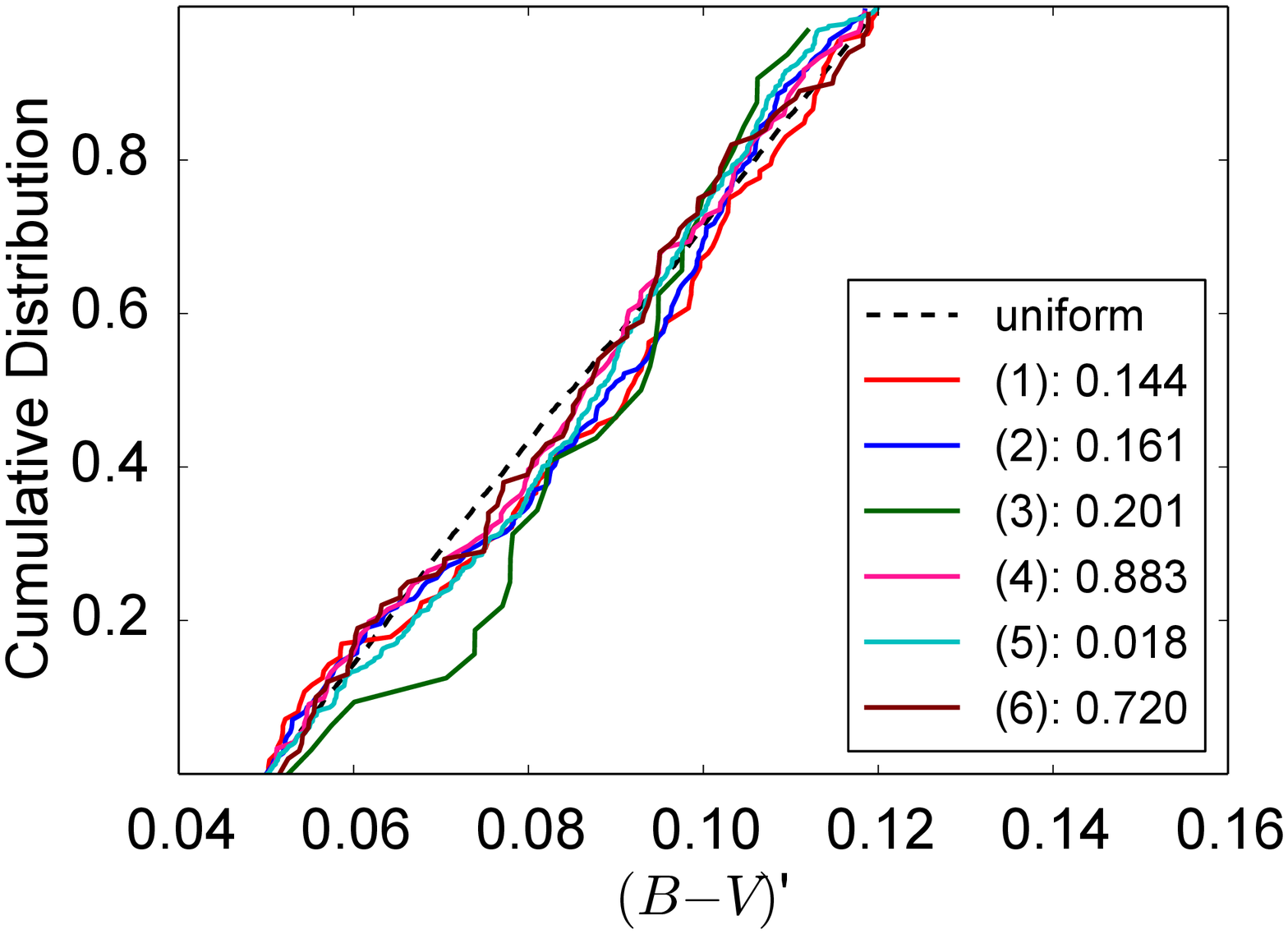} 
    \end{center}
  \caption{Colour distributions derived from the CMDs shown in Fig. \ref{FIG06.1},
following the procedure described in Section \ref{problem_model}. Notations as
in  \ref{FIG03.2}.
Only stars with $(B-V)'>0.05$ are used to compute the above distributions.
Note that in all colour distributions  in the six models following the modified
Kroupa IBP are are consistent uniform distributions. This is supported by the 
statistical test, which does not allows us to reject the hypothesis that they are 
uniform with more than 99 \% of confidence (see $p$-values). 
For more detail see Sections \ref{gccmd}.
}
  \label{FIG06.2}
\end{figure}

After showing that the modified Kroupa IBP provides good agreement
with observational data of Galactic field binaries (i.e. our modifications
do not worsen previously established constraints), we can
turn to the solution of our initial problem.

In Section \ref{problem}, we presented the motivation for our investigation
showing the present-day CMDs and colour distributions derived from them of
a set of six GC models  with different initial and present-day properties. 
All six models there were set with the original Kroupa IBP (Section \ref{original}).
While comparing with observation of the nearest GCs (M4 and NGC6397), we showed
that models set with the original Kroupa IBP are not able to predict realistic binary
distributions in GCs. This is because it provides not only a clearly visible binary 
main-sequence  (due to short-period low-mass binaries with mass ratios equal to 
unity) which is not seen in the CMDs of real GCs, but it also predicts CMD colour
distributions below the turn-off peaked at the right edge of the distributions (associated 
again with high-q binaries). 

We can at this point perform a similar analysis as done in Section \ref{problem},
but taking into account our modified Kroupa IBP (Section \ref{modified}). 
We thus simulate models with the same six initial cluster conditions and perform 
photometry of present-day models. The six model, now set with the modified 
Kroupa IBP, properties are shown in Table 2 and Fig. \ref{FIG06.1} exhibits 
their present-day CMDs.

Note that present-day GC model properties are similar for both set
of models (compare with Table 1), which indicates that our modifications
do not play a key role in the cluster dynamical evolution. However,
while comparing Figs. \ref{FIG06.1} and \ref{FIG03.1} we do notice huge
differences. 

First, that pronounced and clearly visible binary sequence
close and above the turn-off is not present in models set with the modified
Kroupa IBP. Second, we do not clearly see binary sequences below the
turn-off associated with high-q binaries in all models. 
This is a result of our revised pre-main-sequence eigenevolution prescription which
gives a mass-dependent strength of the changes to binary properties (Section 
\ref{method}). This leads to a reduced fraction of high mass ratio binaries 
as well as a uniform spread in the mass ratio in the range between 0.9 and 1.0, 
for those binaries that would have equal mass components.

Proceeding further, we generated colour distribution is the same way
described in Section \ref{problem_model}, which are shown in Fig. \ref{FIG06.2}.
Notice that all histograms seem to be consistent with uniform distributions,
as in observations. Again, we applied one-sample Kolmogorov-Smirnov tests
for uniformity. The $p$-values of these tests are displayed in the bottom
panel of Fig. \ref{FIG06.2}. The tests do not permit us to rule out the
null hypothesis that they are uniform with more than 99 per cent of confidence, 
even though we can reject the null hypothesis for model 5 with
more than 98 per cent of confidence.

We conclude then that our modifications to the Kroupa IBP bring present-day
GC models closer to real GCs, which corresponds to a step forward towards a better
prescription for initial cluster conditions.

We emphasize that the solution to the problem with respect to present-day GC CMDs
(without worsening previous results)
presented here open the possibility that a universal IBP might be at the origin of the binary 
population in both the Galactic field and in GCs.
This is consistent with the conclusions reached by \citet{Leigh_2015} and is important for a 
better understanding of the star formation process in proto-clusters.


\section{CONCLUSIONS AND PERSPECTIVE}
\label{discussion}

The original description of pre-main-sequence eigenevolution of a birth
binary population by \citet{Kroupa_1995b}, while being consistent with
Galactic-field and open cluster binary-star data, is found here to
lead to too many $q=1$ systems which implies that the CMDs of GCs
ought to have pronounced binary sequences above the turn-off and to
the right of the MS. This is not observed to be the case. In this
contribution we revise th Kroupa pre-main-sequence eigenevolution
model such that it is consistent with the Galactic-field, open
cluster and GC data.

To achieve this, we assumed a mass-dependent strength for 
pre-main-sequence eigenevolution such that the lower the primary mass, 
the weaker the changes in the birth population. This procedure results in a 
smaller fraction of short-period binaries with equal mass components which 
provides qualitatively similarly good description of
observations of Galactic field late-type binaries and GCs.

We also assumed distributions for massive binaries such that they
follow directly distributions derived from observations of O-dwarfs in open
stellar clusters. This is consistent with the fact that pre-main-sequence 
timescales are extremely short for massive stars.

Finally, we emphasize that our modified Kroupa IBP should not change results
achieved with the original Kroupa IBP with respect to young star clusters
and GCs. In addition, this paper corresponds to the first step towards a better description
of the IBP that should seed population synthesis codes (after dynamical population
synthesis) and star cluster evolution codes, as well as a better understanding of 
clustered star formation processes such as energy and angular momentum redistribution
within very young binary systems.

In following investigations of the modified Kroupa IBP, 
we will verify the influence of parameters that control
binary stellar evolution such as the energy budget during 
the common-envelope phase and the angular momentum loss formalisms 
for interacting binaries, on populations of white dwarf-main sequence 
post-common-envelope phase binaries and cataclysmic variables. 

\section*{Acknowledgements}

We would like to kindly thank Michael Marks for useful discussions and 
suggestions. We would also like
to thank an anonymous referee for numerous comments and suggestions.
DB was supported by the CAPES foundation, Brazilian Ministry of Education
through the grant BEX 13514/13-0 and by the National Science Centre, Poland,
through the grant UMO-2016/21/N/ST9/02938.
AA would like to acknowledge support from the National Science Centre, Poland, 
through the grant UMO-2015/17/N/ST9/02573 and partial support from Nicolaus
Copernicus Astronomical Centre's grant for young researchers.

\bibliographystyle{mnras}
\bibliography{references}


\appendix

\section[]{Influence of dynamical encounters during stimulated evolution in low-$N$ star clusters}
\label{ap}

Here we show that from the point of view of IBP properties, 
star cluster models with $N=10,000$ objects are reasonable upper limits for 
low-$N$ star cluster models. For this purpose, we will make an
order of magnitude estimate for the period of binaries which 
will certainly interact dynamically (probability of interaction 
equal to 1) with field stars during the time, $\Delta t$.

Let us consider a binary with semi major axis $a$, composed of stars with average stellar 
mass $\langle m \rangle$, which interacts with single stars with average mass, inside the 
half mass radius $R_h$. The probability of interaction is given by 
\citep[e.g.][Eq. 30]{Stodolkiewicz_1986}

\begin{equation}
\Pi \ = \ \pi p_{\rm imp}^2 \eta \upsilon \Delta t,
\label{1}
\end{equation}
\noindent
where $\eta$ is the number density inside the half-mass radius, given by

\begin{equation}
\eta \ = \ \frac{(1/2)(M / \langle m \rangle)}{(4/3)\pi R_h^3},
\label{2}
\end{equation}
\noindent
where $\langle m \rangle = M/N$ (which is constant 
for a given IMF), being $N$ the number of objects (binaries + single stars) and $M$ the star cluster total mass,
$\upsilon$ is the typical relative velocity, which is, assuming an isotropic velocity distribution, given by 

\begin{equation}
\upsilon^2 \ = \ 2 \sigma^2,
\label{3}
\end{equation}
\noindent
where $\sigma$ is the 3D velocity disperstion, given by \citep[][Eq. 1-10]{Spitzer_BOOK}

\begin{equation}
\sigma^2 \ \approx \ 0.4  \, \frac{G M}{R_h},
\label{4}
\end{equation} 
\noindent
$p_{\rm imp}$ is the impact parameter, given by \citep[][Eq. 6-15]{Spitzer_BOOK}

$$p_{\rm imp}^2 \ = \ a^2 \left( 1 + \frac{2G M_{123}}{\upsilon^2 a} \right),$$
\begin{equation}
\Leftrightarrow p_{\rm imp}^2 \ \approx \ 2 \, \frac{G \, M_{123} \, a}{\upsilon^2},
\label{5}
\end{equation}
\noindent
because in embedded clusters, usually $\left( 2G M_{123} \right)/\left(\upsilon^2 a \right) \gg 1$, where
$M_{123}$ is the sum of the masses (binary mass plus single star mass).

Now, replacing the terms in Eq. \ref{1} with Eqs. \ref{2}, \ref{3}, and \ref{5}, 
and assuming that only average stars take part in the interaction, i.e. 
$M_{123} \approx 3 \, \langle m \rangle$, we have

\begin{equation}
\Pi \ \approx \ 1.1 \, \frac{G^{1/2}M^{1/2}}{R_h^{5/2}} \, a \, \Delta t,
\label{6}
\end{equation}
\noindent
which gives, assuming that the star cluster follows the Marks-Kroupa relation 
(Eq. \ref{EQ16}) between $R_h$ and $M$,

\begin{equation}
\Pi \ \approx \ 1.17 \times 10^{-5} \left( \frac{M}{{\rm M}_\odot} \right)^{7/40} 
\left( \frac{a}{{\rm AU}} \right) 
\left( \frac{\Delta t}{{\rm Myr}} \right).
\label{7}
\end{equation}
\noindent
Equating the probability to unity we will get an estimation of the lower 
limit of the binary semimajor axis above which the binary will certainly interact 
with a field star during a time $\Delta t$, i.e. 

\begin{equation}
a \ \approx 8.55 \times 10^{4}  \left( \frac{M}{{\rm M}_\odot} \right)^{-7/40} 
\left( \frac{\Delta t}{{\rm Myr}} \right)^{-1}.
\label{8}
\end{equation}
\noindent
Now, from the Kepler's third law, we have

$$
\left( \frac{P}{{\rm yr}} \right)^{2} \ = \  
\left( \frac{a}{{\rm AU}} \right)^{3}
\left( \frac{M_{12}}{{\rm M}_\odot} \right)^{-1},
$$
\begin{equation}
\left( \frac{P}{{\rm d}} \right) \ \approx \ 9.13 \times 10^{9}
\left( \frac{M}{{\rm M}_\odot} \right)^{-21/80}
\left( \frac{\Delta t}{{\rm Myr}} \right)^{-3/2},
\label{9}
\end{equation}
\noindent
where $M_{12} \approx 2 \langle m \rangle \approx 1 \, {\rm M}_\odot$ is the average binary mass,
and $P$ is the period.

Finally, let us consider three cluster masses, namely $10^2$ M$_\odot$,
$10^3$ M$_\odot$, and $10^4$ M$_\odot$, and estimate the minimum
period for interactions in each of these clusters, during stimulated
evolution ($\Delta t \approx 5$ Myr):
 
\begin{itemize}
\item $M \approx 10^4$ M$_\odot$ $\Leftrightarrow N \approx 10^4$ : $ \ P \ \approx \ 7.3 \times 10^{7} \ {\rm days}$,

\item $M \approx 10^3$ M$_\odot$ $\Leftrightarrow N \approx 10^3$ : $ \ P \ \approx \ 1.3 \times 10^{8} \ {\rm days}$,

\item $M \approx 10^2$ M$_\odot$ $\Leftrightarrow N \approx 10^2$ : $ \ P \ \approx \ 2.4 \times 10^{8} \ {\rm days}$.
\end{itemize}
\noindent

Notice that the smaller the number of objects in the cluster is
(or the cluster mass), the longer is the minimum period
for interactions. This implies that for low-$N$ clusters, mainly 
long-period binaries are affected by dynamical interactions.
Note also that we assumed many simplifications in this calculation,
which leads to an overestimation of such a minimum period. In reality, clusters
are denser in the central parts and composed of objects with different masses, 
which leads to larger probabilities of binary interactions and in turn to smaller
semimajor axis. Said that, the minimum period is likely to be shorter than 
that derived here. However, we do not expect substantial (many orders of magnitude) 
reduction of the estimated minimum period. This means that dynamical interactions 
between binaries and other objects during stimulated evolution will not have any 
influence on the bulk of binaries. Only the long-period tail of the distribution can 
be slightly changed. This leads us to the conclusion that very-low-$N$ clusters are 
dynamically equivalent \citep{Kroupa_1995b} to our cluster model with $N = 10^4$.

\section[]{Dynamical Equivalence}
\label{ap2}

Here we show that the principle of dynamical equivalence associated with
the stimulated evolution is valid. As stated in Section \ref{stimulated}, 
if two clusters with different masses and different initial characteristic radii dynamically 
evolve an identical initial binary population to similar distribution functions 
of binaries after stimulated evolution, then these two clusters are 'dynamically 
equivalent'. In order to show that we can find star cluster models
that are dynamically equivalent, even for different initial conditions, we 
simulated evolution of three different cluster models with the {\sc mocca} code. 
Properties of those models are summarized in Table \ref{TabAP}.

\begin{figure*}
   \begin{center}
    \includegraphics[width=0.48\linewidth]{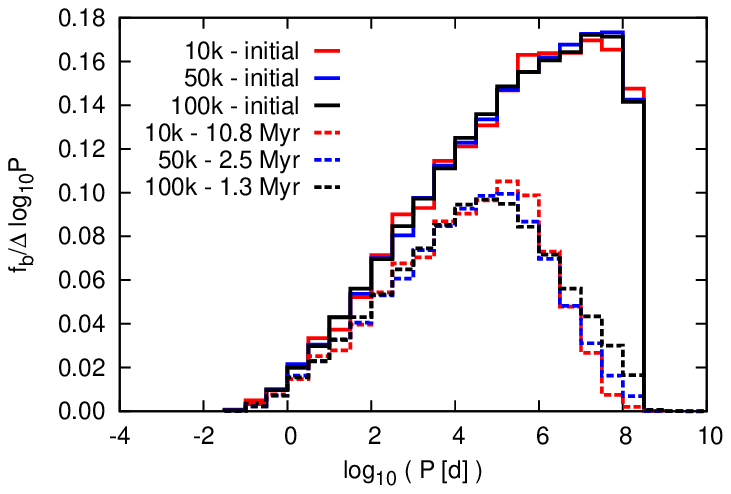} 
    \includegraphics[width=0.48\linewidth]{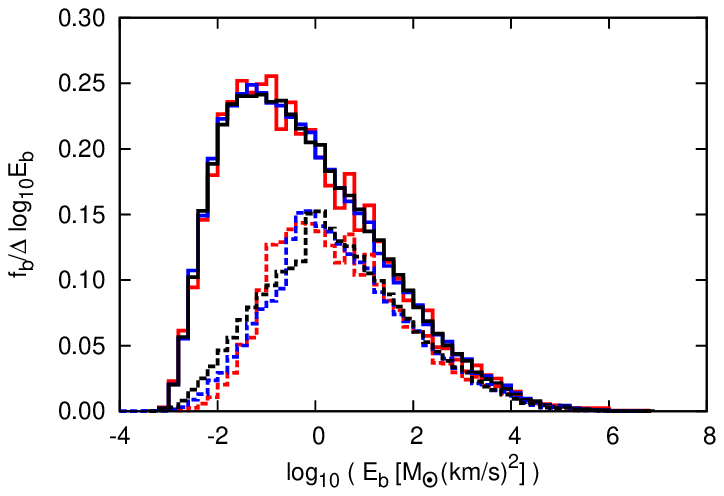} 
    \includegraphics[width=0.48\linewidth]{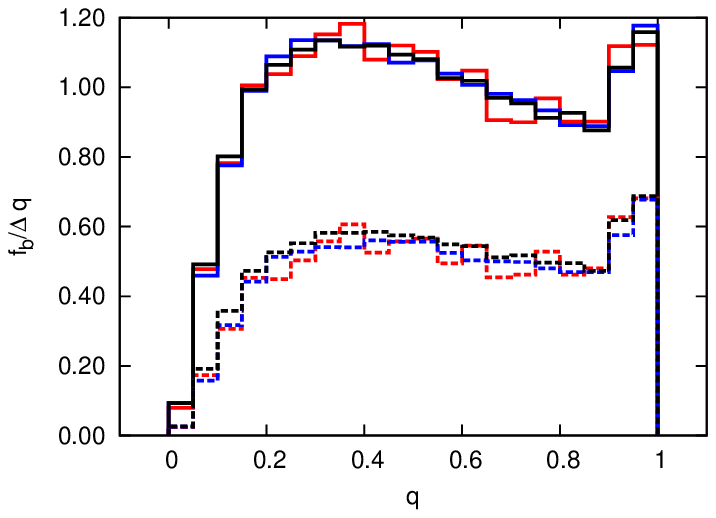} 
    \includegraphics[width=0.48\linewidth]{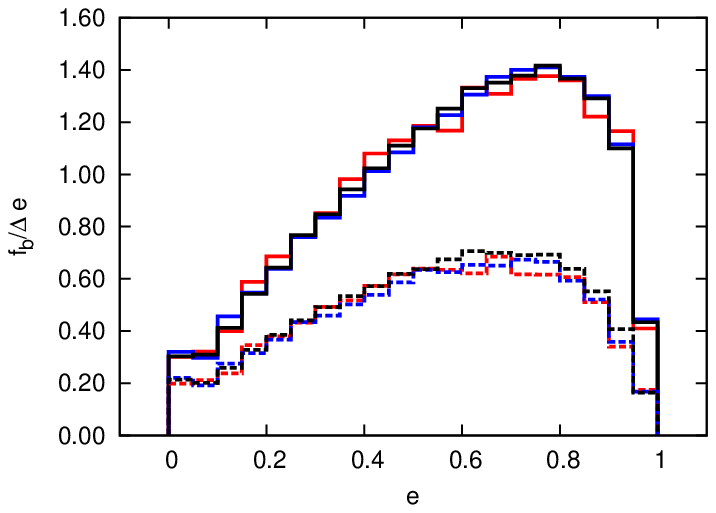} 
    \end{center}
  \caption{Distribution of periods (top left-hand panel), binding energies (top right-hand panel), 
mass ratios (bottom left-hand panel) and eccentricities (bottom right-hand panel) for all binaries.
We depict initial properties as well as properties after stimulated evolution. The area under each 
distribution equals the total binary fraction. Notice that the distributions after stimulated
evolution are remarkably consistent with each other. In addition, each cluster evolutionary
timescale is different, which is needed for dynamical equivalence amongst them, since they
have different initial properties. The prime lesson from this figure is that different
initial cluster conditions can be evolved into similar binary distribution functions,
provided that they evolve during different timescales. For more details see Appendix 
\ref{ap2}.}
  \label{FIG_AP}
\end{figure*}

\begin{table} 
\caption{Initial conditions of the three models used to show that the
principle of dynamical equivalence is valid. As in previous simulations,
for all three models, we assumed solar metallicity and set the initial 
binary fraction as 95 per cent.
The IBP in all models is the modified Kroupa IBP. Notice that the three models 
have very distinct initial conditions, especially
masses and half-mass radius relaxation times.}
\label{TabAP}
\begin{adjustbox}{max width=\linewidth}
\noindent
\begin{threeparttable}
\noindent
\begin{tabular}{l|c|c|c}
\hline\hline
& \multicolumn{3}{c}{Model} \\
\hline
Property & 10k & 50k & 100k \\
\hline
Mass [$\times 10^4 \, {\rm M}_\odot$]					& $0.89$ & $4.60$ & $9.11$	\\
Number of objects [$\times 10^4$] 					& $1$  & $5$ & $10$		\\
Central density	[$\times 10^4 \, {\rm M}_\odot \; {\rm pc}^{-3}$]	& $1.1$ & $2.6$ & $5.3$ 	\\
Tidal radius [pc]							& $25.8$ & $44.1$ & $55.6$	\\
Half-mass radius [pc]		 					& $0.83$ & $1.05$ & $1.14$	\\
Core radius [pc]		 					& $0.28$ & $0.35$ & $0.30$	\\
Galactocentric distance [kpc]	 					& $8.0$ & $7.9$ & $7.9$		\\
Half-mass radius relaxation time [Myr]					& $30.7$  & $75.2$ & $109.2$	\\
Central velocity dispersion [km s$^{-1}$]				& $4.3$ & $11.1$ & $16.1$	\\
Duration of stimulated evolution [Myr]					& $10.8$  & $2.5$ & $1.3$	\\
\hline\hline
\end {tabular}
\end{threeparttable}
\end{adjustbox}
\end{table}

We notice that, since two 
clusters with different initial conditions have different dynamical histories, 
in order to have an equivalent evolution, the timescale of the evolution has to be 
different. Indeed, this is shown in Fig. \ref{FIG_AP} where we depict the period, 
mass ratio, eccentricity and binding energy distributions associated with all 
binaries in the three clusters, initially and after stimulated evolution. 
As it is well known, when we evolve
the three different initial models over the same physical time, their binary distributions
will differ. However, if we control the stimulated evolution timescale, we can
find a value such that their distributions will be similar after that time. 
In the particular case of the three models described in Table \ref{TabAP},
if the timescales for 10k, 50k and 100k models are close to 10.8, 2.5 and 1.3 Myr,
respectively, then all three clusters are dynamically equivalent. In general,
there are pragmatically many combinations of stimulated evolution timescales such 
that these particular three clusters are dynamically equivalent and the combination 
above is just one example.

This simple exercise illustrates the
principle of dynamical equivalence and
its application, i.e. we can
always find dynamical equivalent solutions to the problem of initial
conditions, provided that each cluster evolves at its own particular
timescale.

\bsp

\label{lastpage}

\end{document}